%% file: main.tex
\documentclass[journal=jctcce,manuscript=article,layout=twocolumn,email=false]{achemso}

\usepackage[version=3]{mhchem} 
\usepackage{float}
\usepackage{geometry}
\usepackage{graphicx}
\usepackage{caption}
\usepackage{subcaption}
\usepackage{amsmath}
\usepackage{amssymb}
\usepackage{tabularx}
\usepackage{units}
\usepackage{color}
\usepackage{hyperref}
\usepackage{multirow}
\usepackage{adjustbox}
\usepackage{sectsty}   
\usepackage{fancyhdr}

\author{Abdulgani Annaberdiyev}
\affiliation{Department of Physics, North Carolina State University, Raleigh, North Carolina 27695-8202, USA}

\author{Cody A. Melton}
\affiliation{Department of Physics, North Carolina State University, Raleigh, North Carolina 27695-8202, USA}
\alsoaffiliation{Sandia National Laboratories, Albuquerque, New Mexico 87123, USA}

\author{M. Chandler Bennett}
\affiliation{Department of Physics, North Carolina State University, Raleigh, North Carolina 27695-8202, USA}

\author{Guangming Wang}
\affiliation{Department of Physics, North Carolina State University, Raleigh, North Carolina 27695-8202, USA}

\author{Lubos Mitas}
\affiliation{Department of Physics, North Carolina State University, Raleigh, North Carolina 27695-8202, USA}

\abbreviations{AE,ECP,ccECP,QMC}

\title{Accurate atomic correlation and total energies for correlation consistent effective core potentials}

\sectionfont{\large}   

\begin{document}

{\large SAND2019-11580 O }

\begin{abstract}{\centering}

Very recently, we introduced a set of correlation consistent effective core potentials (ccECPs) constructed within full many-body approaches.  
By employing significantly more accurate 
correlated approaches 
we were able to reach a new level of accuracy for the resulting effective core Hamiltonians. We also strived for simplicity of use and easy transferability
into a variety of electronic structure methods in quantum chemistry and condensed matter physics. 
Here, as a reference for future use, we present exact or nearly-exact total energy calculations for these ccECPs.
The calculations cover H-Kr elements 
and are based on the state-of-the-art configuration interaction (CI), coupled-cluster (CC), and quantum Monte Carlo (QMC) calculations with systematically eliminated/improved errors.
In particular, we carry out full CI/CCSD(T)/CCSDT(Q) calculations with cc-pVnZ with up to n=6 basis sets and we estimate the complete basis set limits. 
Using combinations of these approaches, we achieved an accuracy of $\approx$ 1-10 mHa for K-Zn atoms and $\approx$ 0.1-0.3 mHa for all other elements $-$ within about 1\% or better of the ccECP total correlation energies. 
We also estimate the corresponding kinetic energies within the feasible limit of full CI calculations. In order to provide data for QMC calculations, we include 
fixed-node diffusion Monte Carlo energies for each element that give quantitative insights into the fixed-node biases for single-reference trial wave functions. 
The results offer a clear benchmark for future high accuracy calculations in a broad variety 
of correlated wave function methods such as CI and CC as well is in stochastic approaches such as real space sampling QMC.

{\centering
\includegraphics[width=1.0\linewidth]{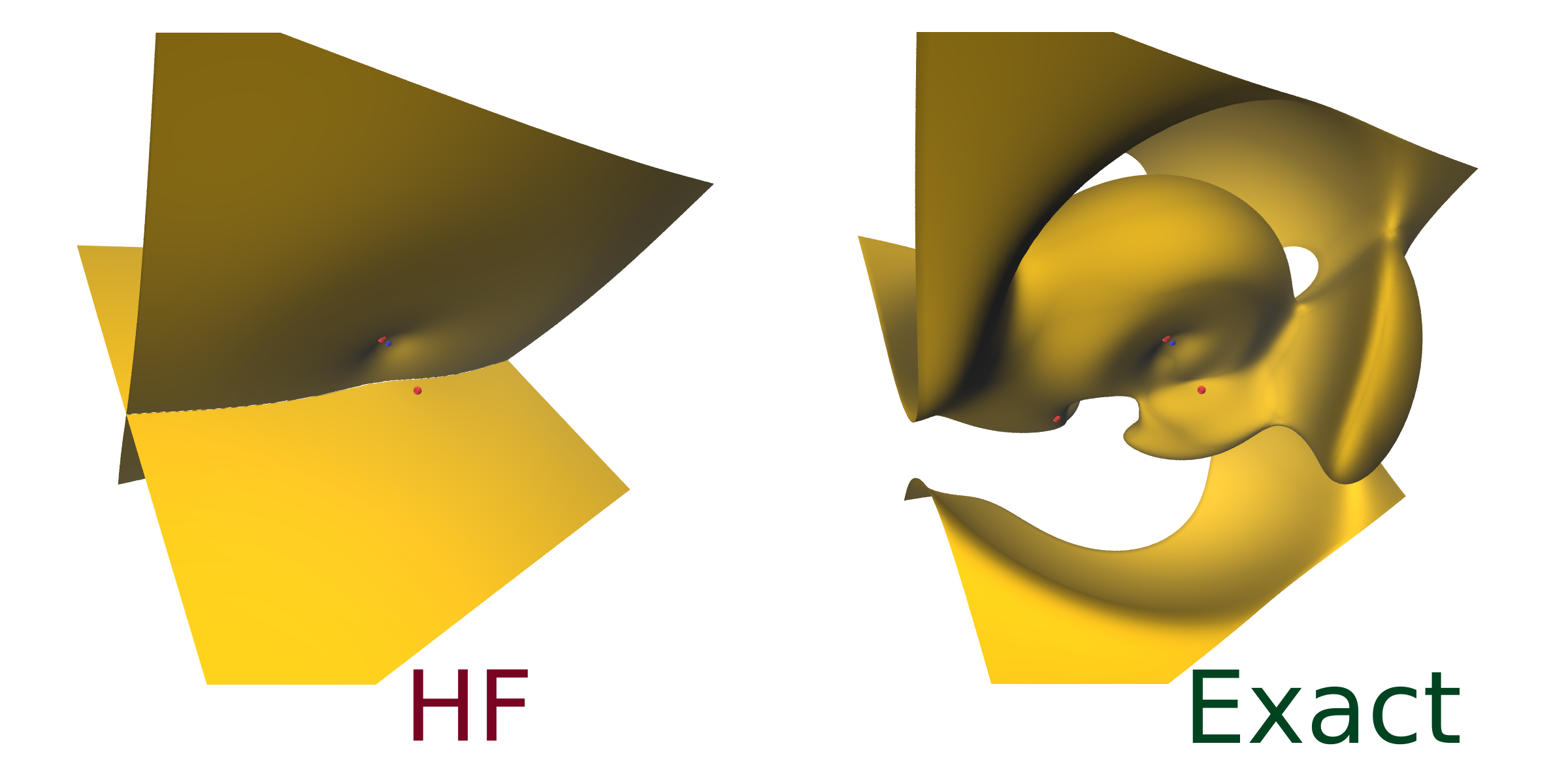}
\par
}

\end{abstract}


\input{intro.tex}

\input{methods.tex}


\input{1st_energy.tex}

\input{2nd_energy.tex}

\input{3rd_energy.tex}


\input{kinetic.tex}


\input{conclusions.tex}

\input{supporting.tex}
\clearpage

\begin{acknowledgement}

We are grateful to Paul R. C. Kent for the reading of the manuscript and helpful suggestions.

This work was supported by the U.S. Department of Energy, Office of Science, Basic Energy Sciences, Materials Sciences and Engineering Division, as part of the Computational Materials Sciences Program and Center for Predictive Simulation of Functional Materials.

This research used resources of the National Energy Research Scientific Computing Center (NERSC), a U.S. Department of Energy Office of Science User Facility operated under Contract No. DE-AC02-05CH11231.

The authors also acknowledge the Texas Advanced Computing Center (TACC) at The University of Texas at Austin for additional HPC resources.

This paper describes objective technical results and analysis. Any subjective views or opinions that might be expressed in the paper do not necessarily represent the views of the U.S. Department of Energy or the United States Government. Sandia National Laboratories is a multimission laboratory managed and operated by National Technology \& Engineering Solutions of Sandia, LLC, a wholly owned subsidiary of Honeywell International Inc., for the U.S. Department of Energy's National Nuclear Security Administration under contract DE-NA0003525.


\end{acknowledgement}

\bibliography{main.bib}

\end{document}

%% file: intro.tex
\section{Introduction}\label{intro}

Effective core potentials (ECPs) or closely related pseudopotentials have been
used very successfully in electronic structure 
calculations for decades. 
ECPs represent a class of effective Hamiltonians that
combine significant gains in computational efficiency with the simplification of describing the valence electronic properties of matter in perhaps the tidiest and most compact formulation.  
ECPs fulfill these multiple roles very successfully and have enabled studies 
that would be otherwise unreachable. For example, smoothing out the ionic potentials and core states has been a simple yet significant step that enabled broad use of plane wave basis sets with major impact in condensed matter physics and in \textit{ab initio} molecular dynamics.   Additional advantages and necessities of using ECPs for very heavy elements have been well-known and demonstrated over decades \cite{dolg-rev}. 

Despite ECP's usefulness, their accuracy has been improving only gradually and sometimes lagged behind the progress 
of methods that treat the fundamental challenge of electron-electron correlations. 
This lag can be understood by the complications  of
new and more accurate constructions that typically required significant time and effort. We believe that this has been the case also in recent years where the electron correlation methods have made significant advances represented, for example, by stochastic configuration interaction approaches, auxiliary field, and real space sampling quantum Monte Carlo (QMC) \cite{Alavi, Zhang, QMCPack}, while employed ECPs were based on more traditional and older construction schemes. 

In order to avoid laborious retesting and revalidation of older pseudopotential sets, 
we have provided a new generation of correlation consistent effective core potentials/pseudopotentials (ccECPs) for the first 36 elements of the periodic table, up to Kr \cite{1-ccECP,2-ccECP,3-ccECP,4-ccECP}.
The key aims of our effort have been the construction of ECPs in a full many-body and scalar relativistic framework, high accuracy, and broad use potential. 
Our construction has been based on
finding the best compromises in reproducing many-body atomic spectra
(i.e, isospectrality for atomic valence states),  molecular binding curves for selected systems,
and high reliability for non-equilibrium molecular geometries (see references \cite{hou_fixed-node_2018, zhou_diffusion_2019} for independent tests).
Our very simple, gaussian parametrization enables straightforward use in majority of codes with almost any basis set type, both in quantum chemical and condensed matter physics applications.
We believe that ccECPs  
offer the community the opportunity to focus on 
physics and chemistry of the given problem rather than on taxing technicalities of the ECP accuracy or construction limitations. In addition,
we have argued that accurate ccECPs, aside from
their use to decrease the degrees of freedom and eliminating the core energy scales,
provide another welcome advantage.
In particular, we have shown that systematic biases from
ccECPs for valence properties are on par or {\it smaller}
than all-electron calculations with uncorrelated cores. 
We achieved this by employing highly accurate correlated constructions of 
ccECPs that took into account also
the core-core and core-valence 
correlations. Remarkably, 
this makes the
resulting effective ECP Hamiltonians  more accurate than most ordinary treatments of cores in all-electron
calculations.
The ccECPs and corresponding basis sets 
up to 6Z has been tabulated and data are readily available \cite{website} and open to further documented updates.

In this work, we present
important reference data that will be useful for subsequent calculations, namely
exact/nearly-exact
ccECP total energies using sequences of highly accurate correlated wave function methods. For H-Kr atoms, we carry out calculations with approaches that include CI single and double excitations (CISD), full CI (FCI), coupled cluster (CC) up to double and triple excitations with perturbative triples 
(CCSD(T)), and limited CC with explicit triples and perturbative quadruples (CCSDT(Q)).
The corresponding basis sets are systematically increased
from cc-pVDZ to cc-pV6Z until we reach the feasibility limits for 
the largest cases.
These finite basis results are then systematically extrapolated 
to the complete basis set (CBS) limits.

Once we have established systematic data on exact/nearly exact total energies our interest includes presenting
insights into recently developed 
selected-CI methods where we have opted for the CIPSI approach 
\cite{CIPSI}.

Our next goal has been to provide data for the fixed-node diffusion Monte Carlo (DMC) studies. In particular, we were interested in revealing both absolute and relative biases that combined fixed-node and localization errors for single-reference trial functions
that were based on HF and Density Functional Theory (DFT) orbitals. 
Further, in order to get an insight into improvements with multi-reference wave functions, we also calculated energies of the transition metal series and a few other elements with trial functions containing up to 
$\approx$ 1 million determinants.

  While in all-electron 
nonrelativistic calculations the kinetic energies are straightforward to obtain,
for ECP
Hamiltonians the kinetic energies are genuinely 
challenging and
require explicitly correlated 
wave functions such as high
level CI. In order to shed a light on this important quantity we  
also calculated the total kinetic energies using the CI method with extrapolations.




The paper is structured as follows. 
Section \ref{methods} provides general information about ccECPs and describes the methods used in this work.
In sections \ref{1st-energies}, \ref{2nd-energies}, and \ref{3rd-energies} we give the estimated exact energies for H-Ne, Na-Ar, and K-Kr elements, respectively. The next section presents CIPSI calculations. 
Section \ref{FNDMC} provides the data obtained in fixed-node DMC calculations. 
Section \ref{kinetic} contains the kinetic energies for the aforementioned set of elements.
Finally, we close with conclusions and discussions in section \ref{conclusion}.

%% file: methods.tex
\section{ccECP form and methods}\label{methods}

The ccECPs have the following well-established semi-local form:
\begin{equation}
    V^{\rm ECP}_i = V_{loc}(r_i) + \sum_{l=0}^{l_{max}} V_l(r_i) \sum_{m}|lm\rangle\langle lm|,
\end{equation}
where $i$ is the electron index, $r_i$ is the electron-ion distance, and $lm$ represents the spherical harmonics. 
Here local potential $V_{loc}(r_i)$ and non-local potential $V_l(r_i)$ are given as follows:
\begin{equation}
    V_{loc}(r) = -\frac{Z_{\rm eff}}{r}(1 -e^{-\alpha r^2}) + \alpha Z_{\rm eff} re^{-\beta r^2} + \sum_{k} \gamma_{k} e^{-\delta_{k} r^2},
\end{equation}

\begin{equation}
    V_l(r) = \sum_{j}\beta_{lj} e^{-\alpha_{lj} r^2},
\end{equation}
where the ranges of $k$ and $j$ sums are typically between 1 and 4.
All greek letters are optimized parameters and  $Z_{\rm eff} = Z - Z_{\rm core}$.
Detailed information on ccECPs with corresponding references can be found at the website \cite{website}.

Throughout the paper, we calculate the atomic energies with various post-HF methods with increasing basis set levels as stated previously. 
We also provide CBS limit values which are obtained separately for HF and correlation energies using an extrapolation scheme \cite{extrapolation} as follows:

\begin{eqnarray}
    \label{eqn:hfextrap}
    E^{\rm HF}_n &=& E^{\rm HF}_{\rm CBS} + a \exp\left[ - b n \right] \\
    \label{eqn:corrextrap}
E^{\rm corr}_{n} &=& E^{\rm corr}_{\rm CBS} + \frac{\alpha}{(n+3/8)^{3}} + \frac{\beta}{(n+3/8)^5}
\end{eqnarray}
where $n$ is the basis set cardinal number. 
The deviations from the fit are interpreted as
standard deviations, i.e, as statistical quantities.
It should be mentioned that SCF energies might show an increase in the 6th or higher digits since the correlation consistent parts of the basis sets were optimized for total energies.
This is not an issue since we constrained the extrapolated SCF CBS values from above by the minimum of obtained SCF energies, and the uncertainty here is much smaller than the uncertainty in correlation energies.

We carry out calculations at $D_{2h}$ point group symmetry as well as using state-averaged (SA) reference states to examine the differences in HF and correlation energies.
State-averaged orbitals are obtained using the MCSCF method by averaging different wave function symmetries (with the same spin) \cite{state-average}.
This procedure provides full atomic symmetry and equivalent, degenerate orbitals expressed with pure spherical harmonics corresponding to the orbital ($s, p, d, ...$).
Unless otherwise specified, all energies shown in this paper correspond to the state-averaged energies, whereas $D_{2h}$ point group calculations are given in the Supporting Information.
For closed-shell atoms such as Ar, Ca and half-full shell atoms such as P, Mn the $D_{2h}$ and SA reference states are equivalent. 

Clearly, for elements with only one valence electron $E^{\rm corr} \equiv 0$ and only
single-electron eigenvalues (denoted as ROHF/RHF) values are given. 
For cases with only two valence electrons, only CCSD is relevant and CISD gives the exact energy for a  given basis set. 
In such cases, only CISD and ROHF values are given.
Similarly, CCSDT(Q) is not relevant for systems with the number of electrons less than or equal to 3.

For the exact total energies, we take the most accurate post-HF method that is feasible for all basis set sizes.
In some cases, the most extensive basis set calculations were not doable and we estimated the energy for those cardinal numbers using an estimate.
In particular, we calculated the ratio of energies obtained from the two most accurate methods with the largest basis set possible.
Then, the missing energies were estimated by multiplying the less accurate method by that ratio. 
This estimator has been partially verified whenever calculations with all the methods were feasible. We have found that the aforementioned ratio changes only very mildly with respect to basis set and gives reasonably consistent estimations.
For instance, the ratio of CCSDT(Q)/UCCSD(T) correlation energies for Oxygen is 1.0022(2), for Sulfur is 1.0056(5), and for Selenium is 1.0048(4) with typical uncertainty given in parentheses.
Usually, FCI energies were used if the number of valence electrons was $\leqslant$ 6 and CCSDT(Q) quantities were used otherwise.

Another set of results is obtained
by the selected-CI  CIPSI \cite{CIPSI} method using the
same basis sets and ccECPs. It offers insights into the performance of this approach for calculations with ECPs.  

For each atom, single-reference fixed-node DMC (FN-DMC) energies were also calculated. 
In addition, for the 3$d$ transition atoms and a few other elements, large
CI expansions were employed as trial wave functions.
All DMC calculations use timestep extrapolation to eliminate error in the Green's function.
We choose timesteps as  $\tau=(0.02, 0.01, 0.005, 0.0025)$ Ha$^{-1}$ and linearly extrapolate to the  zero timestep limit.
In all cases, we employ T-moves algorithm\cite{t-moves} in DMC which evaluates the non-local pseudopotential variationally and gives an upper bound to ground state energy.
Throughout the paper, the single-reference trial wave function multiplied by one-body $J_{eI}$, two-body $J_{ee}$, and three-body $J_{eeI}$ Jastrow factors is used unless otherwise specified.
Single-reference determinants are obtained at $D_{2h}$ point group except for Ti, V, Co, and Ni atoms where state-averaged references were used to obtain a proper single-reference.

The following software packages have been used throughout this paper: \textsc{Molpro}\cite{MOLPRO}, \textsc{Mrcc}\cite{Mrcc}, \textsc{Gamess}\cite{GAMESS}, \textsc{PySCF}\cite{PYSCF},  \textsc{Quantum Package}\cite{QPackage}, \textsc{QWalk}\cite{QWalk} and \textsc{QMCPACK}\cite{QMCPack}.

%% file: 1st_energy.tex
\section{Energies of H-Ne elements}\label{1st-energies}

As mentioned previously, we carried out calculations using both $D_{2h}$ point group symmetry orbitals and state-averaged orbital references. 
We observed that although HF energies and the obtained correlation energies were different in these two cases, the \textit{total} energies were the same within the deviations/errors obtained. 
This is illustrated for oxygen and vanadium ccECP atomic energies in tables \ref{tab:D2h_vs_SA_O} and \ref{tab:D2h_vs_SA_V}, respectively. 
It should be noted that for vanadium, the total energies were different in two cases if CCSD(T) was used (the state-averaged resulted in lower energy). 
However, CCSDT(Q) energies resulted in the same energies within the observed deviations. 
A similar picture was observed for Ni($^3$F) as described below. 
For these reasons, all energies shown in this paper are state-averaged energies unless otherwise specified.

\begin{table}[!htbp]
\setlength{\tabcolsep}{3pt} 
\centering
\caption{Comparison of ccECP atomic energies [Ha] for oxygen obtained with $D_{2h}$ symmetry vs state-averaged (SA)  orbitals. Correlation energies correspond to CCSDT(Q) calculations.
aug-cc-pVnZ basis set.
}
\label{tab:D2h_vs_SA_O}
\begin{adjustbox}{width=\columnwidth,center}
\begin{tabular}{c|cc|cc}
\hline\hline
& \multicolumn{2}{c|}{SA} & \multicolumn{2}{c}{$D_{2h}$} \\
\hline
      &  Corr        &  ROHF           &  Corr        &  ROHF \\
\hline
TZ    & -0.17636278  & -15.68675514  & -0.17347298  & -15.68964442 \\
QZ    & -0.18928076  & -15.68678152  & -0.18628641  & -15.68977530 \\
5Z    & -0.19339178  & -15.68678146  & -0.19039875  & -15.68977400 \\
6Z    & -0.19529685  & -15.68678698  & -0.19229138  & -15.68979202 \\
CBS   & -0.19761(20) & -15.686787(4) & -0.19462(18) & -15.68979(2) \\
\hline
Total & \multicolumn{2}{c|}{-15.88439(20)} & \multicolumn{2}{c}{-15.88441(18)} \\ 
\hline
\hline
\end{tabular}
\end{adjustbox}
\end{table}

\begin{table}[!htbp]
\setlength{\tabcolsep}{3pt} 
\centering
\caption{Comparison of ccECP atomic energies
[Ha] for vanadium obtained with $D_{2h}$ symmetry vs state-averaged (SA)  orbitals.
cc-pCVnZ basis set.
Values with (*) were not feasible to calculate and represent estimates from the calculated data as described in the text.
}
\label{tab:D2h_vs_SA_V}
\begin{adjustbox}{width=\columnwidth,center}
\begin{tabular}{c|cc|cc}
\hline\hline
& \multicolumn{2}{c|}{SA} & \multicolumn{2}{c}{$D_{2h}$} \\
\hline
      &  Corr        &  ROHF            &   Corr        &  ROHF \\
\hline
DZ    & -0.44327182  & -70.89607856   &  -0.45243936  & -70.88650990     \\
TZ    & -0.49636927  & -70.89610942   &  -0.50511842  & -70.88687052     \\
QZ    & -0.52093(*)  & -70.89611611   &  -0.52972(*)  & -70.88703541     \\
5Z    & -0.53130(*)  & -70.89611819   &  -0.54013(*)  & -70.88707243     \\
CBS   & -0.54566(59) & -70.8961185(3) &  -0.55457(58) & -70.88712(3)     \\ 
\hline
Total & \multicolumn{2}{c|}{-71.44178(59)} & \multicolumn{2}{c}{-71.44169(58)} \\ 
\hline
\hline
\end{tabular}
\end{adjustbox}
\end{table}

For the 1st row (period 2), we show ccECP and some selected BFD \cite{BFD-2007,BFD-2008}, eCEPP \cite{TN_ecp} pseudopotential energies with [He] core where we use ccECP[core-size] notation throughout the paper.
For simplicity, we include H and He atoms in this section where these ECPs do not remove any electrons and are used for efficiency purposes.
Table \ref{1,ccECP,SA,en} shows ccECP energies with various methods and basis sets.
In this row, we use aug-cc-pVnZ basis set with n=3-6 to extrapolate to CBS limit using equations \ref{eqn:hfextrap} and \ref{eqn:corrextrap}.
cc-pVnZ basis set was used for Ne as an exception.
For each atom, the highest accuracy methods with all basis cardinal numbers filled were used to calculate the most accurate energies.
For instance, FCI was used in O atom, whereas in F atom, CCSDT(Q) method was used. 
We estimated that the accuracy achieved throughout the 1st row was in the range of $\approx$ 0.1 - 0.3 mHa \footnote{This and other similar accuracy assessments are estimated by considering basis set extrapolation errors as well as the accuracy of the employed methods.}.

\begin{table*}[htbp!]
\setlength{\tabcolsep}{4pt} 
\centering
\small
\caption{Atomic correlation and total energies [Ha] for the 1st-row elements with ccECPs[He] for indicated basis sets and methods.
Post-HF method values correspond to correlation energies.
CBS denotes basis set extrapolated values.  Values with (*) were not feasible to calculate and represent estimates from
the calculated data as described in the text.
aug-cc-pVnZ basis set (cc-pVnZ for Ne).
}
\label{1,ccECP,SA,en}
\begin{adjustbox}{width=0.90\textwidth,center}
\begin{tabular}{l|l|ccccc|r}
\hline\hline
Atom & Method &          DZ &          TZ &          QZ &          5Z &          6Z &             CBS \\
\hline

\multirow{1}{*}{\bf{H}}
& ROHF      & -0.49999965 & -0.49999965 & -0.49999965 & -0.49999965 & &  -0.49999965(1) \\
\hline

\multirow{3}{*}{\bf{He}}
 & CISD  & -0.03390801 & -0.03946245 & -0.04103940 & -0.04156934 & -0.04177354 &    -0.042065(23) \\
 & RHF   & -2.86167947 & -2.86167947 & -2.86167948 & -2.86167948 & -2.86167948 &   -2.86167948(1) \\
\cline{8-8}
 & Total &             &             &             &             &             &    -2.903745(23) \\
\hline

\multirow{1}{*}{\bf{Li}}
& ROHF      & -0.19685279 & -0.19685279 & -0.19685279 & -0.19685279 & & -0.19685279(1)  \\
\hline

\multirow{3}{*}{\bf{Be}}
& CISD  & -0.04699395 & -0.04781761 & -0.04806189 & -0.04818173 & -0.04824394 &   -0.04834636(12) \\
& RHF   & -0.96189258 & -0.96189258 & -0.96189258 & -0.96189259 & -0.96189258 &   -0.96189260(2)  \\
\cline{8-8}
& Total &             &             &             &             &             &   -1.01023896(12) \\
\hline

\multirow{6}{*}{\bf{B}}
& CISD     & -0.06487328 & -0.07156133 & -0.07303256 & -0.07336789 & -0.07358930 &   -0.07372(12) \\
& RCCSD(T) & -0.06599169 & -0.07329245 & -0.07489041 & -0.07525660 & -0.07550263 &   -0.07565(13) \\
& UCCSD(T) & -0.06600063 & -0.07334521 & -0.07494551 & -0.07530921 & -0.07555505 &   -0.07570(14) \\
& FCI      & -0.06639834 & -0.07381139 & -0.07537084 & -0.07571678 & -0.07595057 &   -0.07608(13) \\
& ROHF      & -2.53923039 & -2.53923039 & -2.53923039 & -2.53923045 & -2.53923052 &  -2.5392306(5) \\
\cline{8-8}
& Total    &             &             &             &             &             &   -2.61531(13) \\
\hline

\multirow{7}{*}{\bf{C}}
& CISD     & -0.07858214 & -0.09412068 & -0.09745958 & -0.09827947 & -0.09883435 &   -0.09923(28) \\
& RCCSD(T) & -0.08035936 & -0.09714670 & -0.10075888 & -0.10164492 & -0.10225077 &   -0.10268(31) \\
& UCCSD(T) & -0.08039730 & -0.09728629 & -0.10090187 & -0.10178787 & -0.10239375 &   -0.10282(31) \\
& CCSDT(Q) & -0.08080170 & -0.09784220 & -0.10141577 & -0.10227750 & -0.10286749 &   -0.10327(31) \\
& FCI      & -0.08081223 & -0.09785257 & -0.10142580 & -0.10228723 & -0.10287702 &   -0.10328(31) \\
& ROHF      & -5.31424795 & -5.31424795 & -5.31424795 & -5.31424799 & -5.31424799 &  -5.3142480(1) \\
\cline{8-8}
& Total    &             &             &             &             &             &   -5.41753(31) \\
\hline

\multirow{7}{*}{\bf{N}}
& CISD     & -0.09359382 & -0.11782215 & -0.12363123 & -0.12539949 & -0.12617272 &  -0.127077(65) \\
& RCCSD(T) & -0.09564583 & -0.12165286 & -0.12789000 & -0.12977963 & -0.13061050 &  -0.131573(76) \\
& UCCSD(T) & -0.09570506 & -0.12185906 & -0.12811222 & -0.12999924 & -0.13083009 &  -0.131786(79) \\
& CCSDT(Q) & -0.09598629 & -0.12230085 & -0.12852942 & -0.13038605 & -0.13120254 &  -0.132125(84) \\
& FCI      & -0.09599171 & -0.12230706 & -0.12853580 & -0.13039205 & -0.131208(*) & -0.132131(85) \\
& ROHF      & -9.63386641 & -9.63386651 & -9.63386735 & -9.63386820 & -9.63386789 &  -9.6338682(9) \\
\cline{8-8}
& Total    &             &             &             &             &             &  -9.765999(85) \\
\hline

\multirow{7}{*}{\bf{O}}
& CISD     &  -0.1263632 &  -0.1687295 &  -0.1808646 &  -0.1848001 &  -0.1866118 &   -0.18887(16) \\
& RCCSD(T) &  -0.1305789 &  -0.1757666 &  -0.1886816 &  -0.1928195 &  -0.1947417 &   -0.19709(20) \\
& UCCSD(T) &  -0.1306644 &  -0.1759383 &  -0.1888722 &  -0.1930115 &  -0.1949337 &   -0.19728(20) \\
& CCSDT(Q) &  -0.1309792 &  -0.1763628 &  -0.1892808 &  -0.1933918 &  -0.1952968 &   -0.19761(20) \\
& FCI      &  -0.1309750 &  -0.1763614 &  -0.1892800 &  -0.1933(*) &  -0.1952(*) &   -0.19760(20) \\
& ROHF      & -15.6867449 & -15.6867551 & -15.6867815 & -15.6867815 & -15.6867870 &  -15.686787(4) \\
\cline{8-8}
& Total    &             &             &             &             &             &  -15.88439(20) \\
\hline

\multirow{7}{*}{\bf{F}}
& CISD     &  -0.14676956 &  -0.21409007 &  -0.23553772 &  -0.24226739 &  -0.24595195 &   -0.25002(91) \\
& RCCSD(T) &  -0.15018483 &  -0.22253878 &  -0.24582057 &  -0.25305299 &  -0.25705640 &    -0.2614(10) \\
& UCCSD(T) &  -0.15019866 &  -0.22261665 &  -0.24592715 &  -0.25316670 &  -0.25717430 &    -0.2615(10) \\
& CCSDT(Q) &  -0.15036092 &  -0.22287630 &  -0.24621842 &  -0.25345060 &  -0.2575(*)  &    -0.2618(11) \\
& FCI      &  -0.15037041 &  -0.22287251 &              &              &              &                \\
& ROHF      & -23.93570738 & -23.93582061 & -23.93583615 & -23.93584334 & -23.93584540 &  -23.935848(1) \\
\cline{8-8}
& Total    &              &              &              &              &              &   -24.1976(11) \\
\hline

\multirow{7}{*}{\bf{Ne}}
& CISD     &  -0.18025458 &  -0.24989860 &  -0.29174759 &  -0.30201083 &  -0.30627967 &     -0.30949(91) \\
& RCCSD(T) &  -0.18462256 &  -0.25961485 &  -0.30453033 &  -0.31557498 &  -0.32014022 &     -0.32361(94) \\
& UCCSD(T) &  -0.18462256 &  -0.25961472 &  -0.30453031 &  -0.31557497 &  -0.32014021 &     -0.32361(94) \\
& CCSDT(Q) &  -0.18480612 &  -0.25973913 &  -0.30463859 &  -0.31566443 &  -0.3202(*)  &     -0.32368(96) \\
& FCI      &  -0.1848097 &  -0.2597312 &             &             &             &                       \\
& RHF      & -34.70881857 & -34.70881857 & -34.70881857 & -34.70881857 & -34.70881857 &  -34.70881857(3) \\
\cline{8-8}
& Total    &              &              &              &              &              &    -35.03250(96) \\

\hline\hline
\end{tabular}
\end{adjustbox}
\end{table*}

Table \ref{1,BFD,SA,en} shows energies for selected BFD and eCEPP atoms for comparison of correlation energies with our ccECPs.
We find that the correlation energies in BFD ECPs are typically larger compared to ccECP by values that for few valence electron atoms are of the order of 1-3 mHa ($\approx$ 1 \% of the correlation energy). 
This indirectly points out that ccECP pseudo-orbitals 
are marginally closer to their all-electron 
counterparts decreasing thus the artificial 
increase in correlation energies that has been
identified some time ago \cite{Dolg1996}.
On the other hand, we see that eCEPP and ccECP correlation energies are similar.

\begin{table*}[htbp!]
\setlength{\tabcolsep}{4pt} 
\centering
\small
\caption{Atomic correlation and total energies [Ha] for selected elements with BFD and eCEPP ECPs. Notations as in table \ref{1,ccECP,SA,en}.
aug-cc-pVnZ basis set.
}
\label{1,BFD,SA,en}
\begin{adjustbox}{width=0.80\textwidth,center}
\begin{tabular}{l|l|ccccc|r}
\hline\hline
Atom & Method &          DZ &          TZ &          QZ &          5Z &          6Z &             CBS \\
\hline
\multicolumn{8}{c}{\textbf{BFD}} \\
\hline

\multirow{7}{*}{\bf{C}}
& CISD     & -0.08221503 & -0.09637638 & -0.09940662 & -0.10018296 & -0.10075855 &   -0.10119(30) \\
& RCCSD(T) & -0.08417998 & -0.09953843 & -0.10283250 & -0.10366771 & -0.10429891 &   -0.10477(34) \\
& UCCSD(T) & -0.08422128 & -0.09968033 & -0.10297932 & -0.10381283 & -0.10444425 &   -0.10491(34) \\
& CCSDT(Q) & -0.08466512 & -0.10025500 & -0.10351231 & -0.10431828 & -0.10493279 &   -0.10537(34) \\
& FCI      & -0.08467669 & -0.10026580 & -0.10352289 & -0.10432849 & -0.10494277 &   -0.10538(34) \\
& ROHF      & -5.32895973 & -5.32895973 & -5.32895973 & -5.32895973 & -5.32895974 &  -5.3289597(1) \\
\cline{8-8}
& Total    &             &             &             &             &             &   -5.43434(34) \\
\hline

\multirow{7}{*}{\bf{N}}
& CISD     & -0.09671731 & -0.12005299 & -0.12559499 & -0.12724555 & -0.12802543 &   -0.12888(13) \\
& RCCSD(T) & -0.09894790 & -0.12402643 & -0.12998438 & -0.13174854 & -0.13258862 &   -0.13350(15) \\
& UCCSD(T) & -0.09900824 & -0.12423526 & -0.13020784 & -0.13196952 & -0.13280967 &   -0.13371(15) \\
& CCSDT(Q) & -0.09930931 & -0.12469007 & -0.13063388 & -0.13236535 & -0.13319069 &   -0.13406(15) \\
& FCI      & -0.09931577 & -0.12469623 & -0.13064048 & -0.13237147 & -0.133197(*) &  -0.13407(15) \\
& ROHF      & -9.66837630 & -9.66837630 & -9.66837630 & -9.66837632 & -9.66837636 &  -9.6683764(2) \\
\cline{8-8}
& Total    &             &             &             &             &             &   -9.80244(15) \\
\hline

\multirow{7}{*}{\bf{O}}
& CISD     &  -0.1333741 &  -0.1734360 &  -0.1848289 &  -0.1886201 &  -0.1904063 &   -0.19268(17) \\
& RCCSD(T) &  -0.1381435 &  -0.1809008 &  -0.1930346 &  -0.1970242 &  -0.1989224 &   -0.20135(22) \\
& UCCSD(T) &  -0.1382354 &  -0.1810799 &  -0.1932322 &  -0.1972237 &  -0.1991214 &   -0.20156(22) \\
& CCSDT(Q) &  -0.1385764 &  -0.1815221 &  -0.1936538 &  -0.1976157 &  -0.1994962 &   -0.20182(21) \\
& FCI      &  -0.1385740 &  -0.1815198 &  -0.1936528 &  -0.1976(*) &  -0.1994(*) &   -0.20182(21) \\
& ROHF      & -15.7054257 & -15.7054324 & -15.7054611 & -15.7054612 & -15.7054651 &  -15.705466(4) \\
\cline{8-8}
& Total    &             &             &             &             &             &  -15.90729(21) \\
\hline

\multirow{7}{*}{\bf{Si}}
& CISD     & -0.06973601 & -0.08097778 & -0.08331974 & -0.08395670 & -0.08433717 &   -0.08467(15) \\
& RCCSD(T) & -0.07175392 & -0.08468891 & -0.08744239 & -0.08818864 & -0.08862658 &   -0.08901(16) \\
& UCCSD(T) & -0.07178963 & -0.08492773 & -0.08768571 & -0.08843014 & -0.08886721 &   -0.08925(17) \\
& CCSDT(Q) & -0.07234953 & -0.08582379 & -0.08851680 & -0.08920484 & -0.08961725 &   -0.08994(17) \\
& FCI      & -0.07236952 & -0.08584660 & -0.08854036 & -0.08922708 & -0.08963906 &   -0.08996(17) \\
& ROHF      & -3.67860005 & -3.67859994 & -3.67860002 & -3.67860015 & -3.67860026 &  -3.6786003(3) \\
\cline{8-8}
& Total    &             &             &             &             &             &   -3.76856(17) \\

\hline
\multicolumn{8}{c}{\textbf{eCEPP}} \\
\hline

\multirow{7}{*}{\bf{C}}
& CISD     & -0.0804464 & -0.0949243 & -0.0979623 & -0.0989108 & -0.0992776 &  -0.099750(19) \\
& RCCSD(T) & -0.0823108 & -0.0979892 & -0.1012920 & -0.1023192 & -0.1027136 &  -0.103221(23) \\
& UCCSD(T) & -0.0823505 & -0.0981299 & -0.1014365 & -0.1024632 & -0.1028570 &  -0.103362(23) \\
& CCSDT(Q) & -0.0827820 & -0.0986918 & -0.1019563 & -0.1029529 & -0.1033323 &  -0.103808(20) \\
& FCI      & -0.0827938 & -0.0987023 & -0.1019665 & -0.1029628 & -0.1033420 &  -0.103818(20) \\
& ROHF     & -5.3161624 & -5.3161625 & -5.3161627 & -5.3161633 & -5.3161634 &  -5.3161636(8) \\
\cline{8-8}
&  Total    &            &            &            &            &            &  -5.419981(20) \\
\hline

\multirow{7}{*}{\bf{N}}
& CISD     & -0.0936046 & -0.1179572 & -0.1233717 & -0.1251407 & -0.1258201 &  -0.126758(64) \\
& RCCSD(T) & -0.0956711 & -0.1217856 & -0.1276156 & -0.1295111 & -0.1302327 &  -0.131226(73) \\
& UCCSD(T) & -0.0957296 & -0.1219943 & -0.1278350 & -0.1297297 & -0.1304503 &  -0.131440(72) \\
& CCSDT(Q) & -0.0960162 & -0.1224373 & -0.1282494 & -0.1301146 & -0.1308200 &  -0.131777(69) \\
& FCI      & -0.0960227 & -0.1224433 & -0.1282557 & -0.1301(*) & -0.1308(*) &  -0.131784(69) \\
& ROHF     & -9.6355631 & -9.6355632 & -9.6355633 & -9.6355637 & -9.6355638 &  -9.6355640(6) \\
\cline{8-8}
& Total    &            &            &            &            &            &  -9.767348(69) \\
\hline

\multirow{7}{*}{\bf{O}}
& CISD     &  -0.1275638 &  -0.1703444 &  -0.1816676 &  -0.1855386 &  -0.1871435 &   -0.189430(84) \\
& RCCSD(T) &  -0.1319669 &  -0.1775415 &  -0.1895774 &  -0.1936713 &  -0.1953529 &   -0.197708(88) \\
& UCCSD(T) &  -0.1320501 &  -0.1777175 &  -0.1897704 &  -0.1938667 &  -0.1955470 &   -0.197899(89) \\
& CCSDT(Q) &  -0.1323705 &  -0.1781509 &  -0.1901822 &  -0.1942516 &  -0.1959147 &   -0.198234(89) \\
& FCI      &  -0.1323663 &  -0.1781493 &  -0.1901818 &  -0.1942(*) &  -0.1959(*) &   -0.198233(89) \\
& ROHF     & -15.6488884 & -15.6488755 & -15.6489105 & -15.6489200 & -15.6489245 &   -15.648925(1) \\
\cline{8-8}
& Total    &             &             &             &             &             &  -15.847159(89) \\

\hline\hline
\end{tabular}
\end{adjustbox}
\end{table*}

%% file: 2nd_energy.tex
\section{Energies of Na-Ar elements}\label{2nd-energies}

In the 2nd row (period 3), we calculated ccECP energies for two different core approximations: [Ne] and [He] cores. 
The 2nd row energies are listed for each method and basis in tables \ref{2,ccECP,SA,en} and \ref{2,ccECP-He,SA,en} for [Ne] and [He] cores respectively.  
All extrapolations in the 2nd row used n=3-6 basis cardinal numbers for CBS limit extrapolations. 
For [Ne] core case, we used aug-cc-pVnZ basis set, with Ar being an exception where a basis set without augmentations was used. We estimate that our total energies with [Ne] cores are accurate within $\approx$ 0.1 - 0.2 mHa margins. 
For [He] core calculations, we carried out calculations with cc-pVnZ basis sets.
For 2nd row energies with ccECP[He], we estimate the uncertainty of about 1 mHa for Na and up to approximately 10 mHa for Ar.
Note that the correlation energies in the 2nd row are smaller than their 1st row isoelectronic counterparts reflecting less localized electron densities in the 2nd row. 

\begin{table*}[htbp!]
\setlength{\tabcolsep}{4pt} 
\centering
\small
\caption{Atomic correlation and total energies [Ha] for 2nd-row elements with ccECPs[Ne]. Notation as in table \ref{1,ccECP,SA,en}.
aug-cc-pVnZ basis set (cc-pVnZ for Ar).
}
\label{2,ccECP,SA,en}
\begin{adjustbox}{width=0.95\textwidth,center}
\begin{tabular}{l|l|rrrrr|r}
\hline\hline
Atom & Method &          DZ &          TZ &          QZ &          5Z &          6Z &             CBS \\
\hline

\multirow{1}{*}{\bf{Na}}
& ROHF      &  -0.18583098 & -0.18615968 & -0.18620499 & -0.18620544 & & -0.1862059(2) \\
\hline

\multirow{3}{*}{\bf{Mg}}
& CISD     & -0.03375431 & -0.03488593 & -0.03495923 & -0.03498970 & -0.03503476 &  -0.035077(30) \\
& RHF      & -0.78825768 & -0.78835857 & -0.78839186 & -0.78839376 & -0.78839489 &   -0.788396(3) \\
\cline{8-8}
& Total    &             &             &             &             &             &  -0.823473(30) \\
\hline

\multirow{6}{*}{\bf{Al}}
& CISD     & -0.05266598 & -0.05657464 & -0.05754613 & -0.05785277 & -0.05799045 &   -0.058158(11) \\
& RCCSD(T) & -0.05379510 & -0.05822615 & -0.05932258 & -0.05967092 & -0.05982655 &   -0.060018(12) \\
& UCCSD(T) & -0.05381025 & -0.05829684 & -0.05939321 & -0.05974132 & -0.05989633 &   -0.060087(11) \\
& FCI      & -0.05426781 & -0.05882587 & -0.05986847 & -0.06019660 & -0.06033957 &  -0.0605153(82) \\
& ROHF      & -1.87700762 & -1.87700762 & -1.87700762 & -1.87700763 & -1.87700764 &   -1.8770076(4) \\
\cline{8-8}
& Total    &             &             &             &             &             &   -1.9375230(82) \\
\hline

\multirow{7}{*}{\bf{Si}}
& CISD     & -0.07150678 & -0.08069788 & -0.08289393 & -0.08364535 & -0.08392281 &  -0.084338(48) \\
& RCCSD(T) & -0.07361325 & -0.08438514 & -0.08696645 & -0.08784026 & -0.08816394 &  -0.088641(52) \\
& UCCSD(T) & -0.07365151 & -0.08462369 & -0.08720748 & -0.08808063 & -0.08840297 &  -0.088878(53) \\
& CCSDT(Q) & -0.07422777 & -0.08551343 & -0.08801955 & -0.08884899 & -0.08914322 &  -0.089575(57) \\
& FCI      & -0.07424762 & -0.08553587 & -0.08804238 & -0.08887095 & -0.08916469 &  -0.089595(57) \\
& ROHF      & -3.67247773 & -3.67247773 & -3.67247773 & -3.67247774 & -3.67247776 &  -3.6724778(1) \\
\cline{8-8}
& Total    &             &             &             &             &             &  -3.762073(57) \\
\hline

\multirow{7}{*}{\bf{P}}
& CISD     & -0.0872674 & -0.1041457 & -0.1081973 & -0.1095219 & -0.1100142 &  -0.110706(64) \\
& RCCSD(T) & -0.0899578 & -0.1098976 & -0.1147390 & -0.1162992 & -0.1168722 &  -0.117667(76) \\
& UCCSD(T) & -0.0900033 & -0.1102977 & -0.1151428 & -0.1167031 & -0.1172737 &  -0.118066(78) \\
& CCSDT(Q) & -0.0905108 & -0.1113014 & -0.1160981 & -0.1175918 & -0.1181241 &  -0.118837(74) \\
& FCI      & -0.0905450 & -0.1113280 & -0.1161277 & -0.1176(*) & -0.1181(*) &  -0.118867(74) \\
& ROHF      & -6.3409664 & -6.3409664 & -6.3409664 & -6.3409664 & -6.3409664 &  -6.3409664(1) \\
\cline{8-8}
& Total    &            &            &            &            &            &  -6.459833(74) \\
\hline

\multirow{7}{*}{\bf{S}}
& CISD     & -0.11573448 & -0.14989015 & -0.15937908 & -0.16235815 & -0.16349147 &  -0.164977(84) \\
& RCCSD(T) & -0.12132342 & -0.16074490 & -0.17165174 & -0.17508365 & -0.17636499 &   -0.17807(12) \\
& UCCSD(T) & -0.12139641 & -0.16100792 & -0.17196728 & -0.17540478 & -0.17668674 &   -0.17838(12) \\
& CCSDT(Q) & -0.12200559 & -0.16200899 & -0.17299950 & -0.17637533 & -0.17761235 &   -0.17921(12) \\
& FCI      & -0.12203637 & -0.16204172 & -0.17303666 & -0.17641(*) & -0.17765(*) &   -0.17925(12) \\
& ROHF      & -9.91821904 & -9.91821904 & -9.91821904 & -9.91821905 & -9.91821913 &  -9.9182192(2) \\
\cline{8-8}
& Total    &             &             &             &             &             &  -10.09747(12) \\
\hline

\multirow{7}{*}{\bf{Cl}}
& CISD     &  -0.14344470 &  -0.19210223 &  -0.20806827 &  -0.21302791 &  -0.21493425 &    -0.21738(11) \\
& RCCSD(T) &  -0.15096047 &  -0.20727126 &  -0.22552938 &  -0.23121346 &  -0.23335316 &    -0.23614(17) \\
& UCCSD(T) &  -0.15102002 &  -0.20742042 &  -0.22573339 &  -0.23142436 &  -0.23356478 &    -0.23634(17) \\
& CCSDT(Q) &  -0.15158951 &  -0.20834100 &  -0.22672192 &  -0.23236924 &  -0.23451(*) &    -0.23724(17) \\
& FCI      &  -0.15161123 &  -0.20837119 &              &              &              &                 \\
& ROHF      & -14.68946978 & -14.68946978 & -14.68946978 & -14.68946981 & -14.68946985 &  -14.6894699(2) \\
\cline{8-8}
& Total    &              &              &              &              &              &   -14.92671(17) \\
\hline

\multirow{7}{*}{\bf{Ar}}
& CISD     &  -0.14172846 &  -0.22523355 &  -0.25200143 &  -0.26029839 &  -0.26354799 &     -0.26766(11) \\
& RCCSD(T) &  -0.14635485 &  -0.24208710 &  -0.27329000 &  -0.28294884 &  -0.28664854 &     -0.29138(21) \\
& UCCSD(T) &  -0.14635487 &  -0.24208709 &  -0.27329003 &  -0.28294886 &  -0.28664854 &     -0.29138(21) \\
& CCSDT(Q) &  -0.14660047 &  -0.24266875 &  -0.27404277 &  -0.28371164 &  -0.28736669 &     -0.29204(24) \\
& FCI      &  -0.14661423 &  -0.24268224 &              &              &              &                  \\
& RHF      & -20.77966277 & -20.77966277 & -20.77966277 & -20.77966277 & -20.77966277 &  -20.77966277(1) \\
\cline{8-8}
& Total    &              &              &              &              &              &    -21.07170(24) \\

\hline\hline
\end{tabular}
\end{adjustbox}
\end{table*}

\begin{table*}[htbp!]
\setlength{\tabcolsep}{4pt} 
\centering
\small
\caption{Atomic correlation and total energies [Ha] for 2nd-row elements with ccECPs[He]. Notation same as in table \ref{1,ccECP,SA,en}.
cc-pCVnZ basis set.
}
\label{2,ccECP-He,SA,en}
\begin{adjustbox}{width=0.95\textwidth,center}
\begin{tabular}{l|l|rrrrr|r}
\hline\hline
Atom & Method &          DZ &          TZ &          QZ &          5Z &          6Z &             CBS \\
\hline

\multirow{6}{*}{\bf{Na}}
& CISD     &  -0.18031248 &  -0.27172919 &  -0.29544389 &  -0.30406634 &  -0.30788932 &   -0.313410(61) \\
& RCCSD(T) &  -0.18344571 &  -0.28002589 &  -0.30494689 &  -0.31397705 &  -0.31797296 &   -0.323729(64) \\
& UCCSD(T) &  -0.18344870 &  -0.28003157 &  -0.30495305 &  -0.31398332 &  -0.31797928 &   -0.323736(64) \\
& CCSDT(Q) &  -0.18353692 &  -0.28010235 &  -0.3050(*)  &  -0.3141(*)  &  -0.3181(*)  &   -0.323818(64) \\
& ROHF      & -47.35715946 & -47.35715947 & -47.35715947 & -47.35715948 & -47.35715948 &  -47.3571595(1) \\
\cline{8-8}
& Total    &              &              &              &              &              &  -47.680977(64) \\
\hline

\multirow{6}{*}{\bf{Mg}}
& CISD     &  -0.19418640 &  -0.29280864 &  -0.32236519 &  -0.33189329 &  -0.33599570 &    -0.34124(12) \\
& RCCSD(T) &  -0.20421196 &  -0.30991836 &  -0.34168720 &  -0.35180972 &  -0.35613221 &    -0.36159(13) \\
& UCCSD(T) &  -0.20421197 &  -0.30991843 &  -0.34168729 &  -0.35180982 &  -0.35613232 &    -0.36159(13) \\
& CCSDT(Q) &  -0.20425786 &  -0.31002264 &  -0.3418(*)  &  -0.3519(*)  &  -0.3563(*)  &    -0.36171(13) \\
& RHF      & -62.92742515 & -62.92742515 & -62.92742515 & -62.92742519 & -62.92742527 &  -62.9274253(1) \\
\cline{8-8}
& Total    &              &              &              &              &              &   -63.28914(13) \\
\hline

\multirow{6}{*}{\bf{Al}}
& CISD     &  -0.22008096 &  -0.31527313 &  -0.34539432 &  -0.35648217 &  -0.36115369 &    -0.36818(35) \\
& RCCSD(T) &  -0.23232609 &  -0.33529227 &  -0.36764194 &  -0.37945835 &  -0.38441472 &    -0.39183(37) \\
& UCCSD(T) &  -0.23233690 &  -0.33536902 &  -0.36773002 &  -0.37954815 &  -0.38450517 &    -0.39192(37) \\
& CCSDT(Q) &  -0.23275504 &  -0.33590005 &  -0.3683(*)  &  -0.3801(*)  &  -0.3851(*)  &    -0.39255(37) \\
& ROHF      & -80.99324173 & -80.99324173 & -80.99324173 & -80.99324176 & -80.99324189 &  -80.9932419(2) \\
\cline{8-8}
& Total    &              &              &              &              &              &   -81.38579(37) \\
\hline

\multirow{6}{*}{\bf{Si}}
& CISD     &   -0.2380906  &   -0.3417742  &   -0.3738566  &   -0.3852611  &   -0.3903254  &    -0.3974645(68) \\
& RCCSD(T) &   -0.2507506  &   -0.3641841  &   -0.3988569  &   -0.4110569  &   -0.4164440  &    -0.4239794(44) \\
& UCCSD(T) &   -0.2507694  &   -0.3644072  &   -0.3991011  &   -0.4113022  &   -0.4166890  &    -0.4242204(35) \\
& CCSDT(Q) &   -0.2512525  &   -0.3652314  &   -0.4000(*)  &   -0.4122(*)  &   -0.4176(*)  &    -0.4251799(35) \\
& ROHF      & -101.6260795  & -101.6260795  & -101.6260796  & -101.6260796  & -101.6260798  &   -101.6260798(2) \\
\cline{8-8}
& Total    &               &               &               &               &               &  -102.0512597(35) \\
\hline

\multirow{6}{*}{\bf{P}}
& CISD     &   -0.2543832  &   -0.3690643  &   -0.4034612  &   -0.4157952  &   -0.4213758  &    -0.429233(66) \\
& RCCSD(T) &   -0.2660361  &   -0.3925322  &   -0.4299939  &   -0.4432510  &   -0.4492035  &    -0.457505(72) \\
& UCCSD(T) &   -0.2660517  &   -0.3928744  &   -0.4303578  &   -0.4436120  &   -0.4495627  &    -0.457855(74) \\
& CCSDT(Q) &   -0.2664275  &   -0.3937591  &   -0.4313(*)  &   -0.4446(*)  &   -0.4506(*)  &    -0.458886(74) \\
& ROHF     & -125.2587059  & -125.2587059  & -125.2587059  & -125.2587065  & -125.2587061  &  -125.2587067(5) \\
\cline{8-8}
& Total    &               &               &               &               &               &  -125.717593(74) \\
\hline

\multirow{6}{*}{\bf{S}}
& CISD     &   -0.2798785  &   -0.4143619  &   -0.4545884  &   -0.4690253  &   -0.4754336  &    -0.484560(47) \\
& RCCSD(T) &   -0.2936823  &   -0.4442865  &   -0.4888893  &   -0.5046241  &   -0.5115011  &    -0.521202(86) \\
& UCCSD(T) &   -0.2937266  &   -0.4445279  &   -0.4891849  &   -0.5049308  &   -0.5118094  &    -0.521509(87) \\
& CCSDT(Q) &   -0.2941362  &   -0.4453863  &   -0.4901(*)  &   -0.5059(*)  &   -0.5128(*)  &    -0.522516(87) \\
& ROHF     & -151.9168140  & -151.9168140  & -151.9168140  & -151.9168145  & -151.9168144  &  -151.9168146(5) \\
\cline{8-8}
& Total    &               &               &               &               &               &  -152.439331(87) \\
\hline

\multirow{6}{*}{\bf{Cl}}
& CISD     &   -0.3058358  &   -0.4585030  &   -0.5054814  &   -0.5221170  &   -0.5297420  &     -0.54027(24) \\
& RCCSD(T) &   -0.3206245  &   -0.4926510  &   -0.5452621  &   -0.5635401  &   -0.5717696  &     -0.58301(21) \\
& UCCSD(T) &   -0.3206615  &   -0.4927936  &   -0.5454559  &   -0.5637460  &   -0.5719776  &     -0.58321(21) \\
& CCSDT(Q) &   -0.3210282  &   -0.4935496  &   -0.5463(*)  &   -0.5646(*)  &   -0.57286(*) &     -0.58411(21) \\
& ROHF     & -181.6130087  & -181.6130087  & -181.6130087  & -181.6130093  & -181.6130096  &  -181.6130097(6) \\
\cline{8-8}
& Total    &               &               &               &               &               &   -182.19712(21) \\
\hline

\multirow{6}{*}{\bf{Ar}}
& CISD     &   -0.3291259  &   -0.4983511  &   -0.5527181  &   -0.5720171  &   -0.5808150  &     -0.59302(22) \\
& RCCSD(T) &   -0.3438517  &   -0.5346209  &   -0.5956646  &   -0.6169443  &   -0.6264745  &     -0.63957(18) \\
& UCCSD(T) &   -0.3438517  &   -0.5346209  &   -0.5956646  &   -0.6169443  &   -0.6264745  &     -0.63957(18) \\
& CCSDT(Q) &   -0.3441534  &   -0.5351650  &   -0.5963(*)  &   -0.6176(*)  &   -0.6271(*)  &     -0.64022(18) \\
& RHF      & -214.8921598  & -214.8921598  & -214.8921599  & -214.8921604  & -214.8921611  &  -214.8921612(8) \\
\cline{8-8}
& Total    &               &               &               &               &               &   -215.53238(18) \\

\hline\hline
\end{tabular}
\end{adjustbox}
\end{table*}

%% file: 3rd_energy.tex
\section{Energies of K-Kr  elements}\label{3rd-energies}

The considered 3rd row elements (period 4) consist of K-Kr atoms.
We use [Ne] core ccECP for K, Ca, and $3d$ transition metal (TM) elements while [[Ar]$3d^{10}$] core was used for Ga-Kr.
Here, cc-pCVnZ basis sets (with semicore correlating exponents) were used for the K-Zn elements and aug-cc-pVnZ were used for Ga-Kr. 
Extrapolations for TMs were carried out using n=2-5 since that showed rather systematic data for extrapolations.
Meanwhile for K, Ca and Ga-Kr, n=3-6 basis sizes were used.
Detailed information about 3rd row accurate energies are given in tables \ref{3a,ccECP,SA,en}, \ref{3b,ccECP,SA,en}, and \ref{3rd-main,ccECP,SA,en} which provide the data for Sc-Mn, Fe-Zn and 3rd row main group elements respectively.
In addition, we provide selected coupled-cluster energies for Stuttgart/Cologne group ECPs\cite{STU-1987} (STU)  in Table \ref{3,STU,SA,en} for comparison.

Achieving a high accuracy in K-Zn elements proved to be more challenging due to the significant growth in both total and correlation energies as they stem from
semicore $3s, 3p$ states and from much higher correlations in the $3d-$shell.
We estimate that the uncertainty for the totals throughout K-Zn varies between $\approx$ 1 mHa for the lightest elements to about 10 mHa for Zn.

For Ga-Kr, we estimate an accuracy of 0.1-0.3 mHa. 
Note that these atoms are isoelectronic to $2p$ and $3p$ elements since $d$ electrons are removed by the ECP.
We observe that the correlation energies are again systematically smaller when compared to the previous row ($3p$), suggesting more spread out densities.

\begin{table*}[htbp!]
\setlength{\tabcolsep}{4pt} 
\centering
\small
\caption{Atomic correlation and total energies [Ha] for Sc-Mn with ccECPs[Ne]. Notation as in table \ref{1,ccECP,SA,en}. cc-pCVnZ basis set.}
\label{3a,ccECP,SA,en}
\begin{tabular}{l|l|rrrr|r}
\hline\hline
Atom & Method &          DZ &          TZ &          QZ &          5Z &             CBS \\
\hline

\multirow{6}{*}{\bf{Sc}($^2$D)}
& CISD     &  -0.32941404 &  -0.36328125 &  -0.37550997 &  -0.38166591 &   -0.38834(77) \\
& RCCSD(T) &  -0.36785022 &  -0.40690712 &  -0.42073880 &  -0.42759337 &   -0.43505(81) \\
& UCCSD(T) &  -0.36799510 &  -0.40708064 &  -0.42091692 &  -0.42777322 &   -0.43523(81) \\
& CCSDT(Q) &  -0.36945952 &  -0.40857489 &  -0.42246(*) &  -0.42934(*) &   -0.43684(81) \\
& ROHF      & -46.12000036 & -46.12018898 & -46.12019471 & -46.12019488 &  -46.120198(5) \\
\cline{7-7}
& Total    &              &              &              &              &  -46.55704(81) \\
\hline

\multirow{6}{*}{\bf{Ti}($^3$F)}
& CISD     &  -0.35905310 &  -0.40055231 &  -0.41647131 &  -0.42420355 &   -0.43307(68) \\
& RCCSD(T) &  -0.39968372 &  -0.44707403 &  -0.46501841 &  -0.47370319 &   -0.48363(76) \\
& UCCSD(T) &  -0.39989737 &  -0.44733296 &  -0.46528913 &  -0.47398132 &   -0.48392(77) \\
& CCSDT(Q) &  -0.40136726 &  -0.44879873 &  -0.46681(*) &  -0.47553(*) &   -0.48552(77) \\
& ROHF      & -57.60683700 & -57.60710069 & -57.60711040 & -57.60711048 &  -57.607112(2) \\
\cline{7-7}
& Total    &              &              &              &              &  -58.09263(76) \\
\hline

\multirow{6}{*}{\bf{V}($^4$F)}
& CISD     &  -0.39802486 &  -0.44442781 &  -0.46572515 &  -0.47478335 &    -0.48721(44) \\
& RCCSD(T) &  -0.44158688 &  -0.49458744 &  -0.51906790 &  -0.52939542 &    -0.54369(59) \\
& UCCSD(T) &  -0.44189227 &  -0.49495757 &  -0.51945809 &  -0.52979347 &    -0.54410(59) \\
& CCSDT(Q) &  -0.44327182 &  -0.49636927 &  -0.52093(*) &  -0.53130(*) &    -0.54566(59) \\
& ROHF      & -70.89607856 & -70.89610942 & -70.89611611 & -70.89611819 &  -70.8961185(3) \\
\cline{7-7}
& Total    &              &              &              &              &   -71.44178(59) \\
\hline

\multirow{6}{*}{\bf{Cr}($^7$S)}
& CISD     &  -0.42723842 &  -0.48776726 &  -0.51425496 &  -0.52577647 &   -0.54103(24) \\
& RCCSD(T) &  -0.46229170 &  -0.53117291 &  -0.56131191 &  -0.57436463 &   -0.59170(32) \\
& UCCSD(T) &  -0.46257604 &  -0.53151639 &  -0.56168036 &  -0.57474085 &   -0.59209(33) \\
& CCSDT(Q) &  -0.46308970 &  -0.53189219 &  -0.56207(*) &  -0.57514(*) &   -0.59253(33) \\
& ROHF      & -86.04808378 & -86.04855303 & -86.04855377 & -86.04855525 &   -86.04856(1) \\
\cline{7-7}
& Total    &              &              &              &              &  -86.64109(33) \\
\hline

\multirow{6}{*}{\bf{Mn}($^6$S)}
& CISD     &   -0.4573916  &   -0.5221195  &   -0.5522082  &   -0.5647893  &    -0.58237(83) \\
& RCCSD(T) &   -0.5044855  &   -0.5779711  &   -0.6121330  &   -0.6263267  &     -0.6463(10) \\
& UCCSD(T) &   -0.5049472  &   -0.5784922  &   -0.6126726  &   -0.6268722  &     -0.6468(10) \\
& CCSDT(Q) &   -0.5056657  &   -0.5790926  &   -0.61330(*) &   -0.62752(*) &     -0.6475(10) \\
& ROHF     & -103.2441380  & -103.2443328  & -103.2443341  & -103.2443426  &  -103.244343(6) \\
\cline{7-7}
& Total    &               &               &               &               &   -103.8919(10) \\

\hline\hline
\end{tabular}
\end{table*}

\begin{table*}[htbp!]
\setlength{\tabcolsep}{4pt} 
\centering
\small
\caption{Atomic correlation and total energies [Ha] for Fe-Zn with ccECPs[Ne]. Notation as in table \ref{1,ccECP,SA,en}. cc-pCVnZ basis set.}
\label{3b,ccECP,SA,en}
\begin{tabular}{l|l|rrrr|r}
\hline\hline
Atom & Method &          DZ &          TZ &          QZ &          5Z &             CBS \\
\hline

\multirow{6}{*}{\bf{Fe}($^5$D)}
& CISD     &   -0.5138526  &   -0.5919750  &   -0.6291293  &   -0.6452264  &    -0.66723(65) \\
& RCCSD(T) &   -0.5694493  &   -0.6582469  &   -0.7009070  &   -0.7192144  &    -0.74452(91) \\
& UCCSD(T) &   -0.5699506  &   -0.6588094  &   -0.7015026  &   -0.7198246  &    -0.74515(91) \\
& CCSDT(Q) &   -0.5706896  &   -0.6593646  &   -0.70209(*) &   -0.72043(*) &    -0.74581(93) \\
& ROHF     & -122.6416371  & -122.6422208  & -122.6422246  & -122.6422256  &   -122.64224(2) \\
\cline{7-7}
& Total    &               &               &               &               &  -123.38804(93) \\
\hline

\multirow{6}{*}{\bf{Co}($^4$F)}
& CISD     &   -0.5583647  &   -0.6499413  &   -0.6945658  &   -0.7141035  &    -0.74078(70) \\
& RCCSD(T) &   -0.6178712  &   -0.7216610  &   -0.7729719  &   -0.7952634  &     -0.8260(10) \\
& UCCSD(T) &   -0.6182409  &   -0.7221030  &   -0.7734492  &   -0.7957585  &     -0.8266(10) \\
& CCSDT(Q) &   -0.6186403  &   -0.7222824  &   -0.77364(*) &   -0.79595(*) &     -0.8268(10) \\
& ROHF     & -144.3268501  & -144.3273307  & -144.3273366  & -144.3273370  &  -144.327341(9) \\
\cline{7-7}
& Total    &               &               &               &               &   -145.1541(10) \\
\hline

\multirow{6}{*}{\bf{Ni}($^3$F)}
& CISD     &   -0.6096138  &   -0.7142496  &   -0.7665050  &   -0.7894441  &    -0.82094(87) \\
& RCCSD(T) &   -0.6745776  &   -0.7928604  &   -0.8530765  &   -0.8793837  &     -0.9159(12) \\
& UCCSD(T) &   -0.6748064  &   -0.7931516  &   -0.8533970  &   -0.8797162  &     -0.9162(12) \\
& CCSDT(Q) &   -0.6747698  &   -0.7928519  &   -0.85307(*) &   -0.87938(*) &     -0.9159(12) \\
& ROHF     & -168.4750742  & -168.4752670  & -168.4752698  & -168.4752702  &  -168.475273(4) \\
\cline{7-7}
& Total    &               &               &               &               &   -169.3912(12) \\
\hline

\multirow{6}{*}{\bf{Ni}($^3$D)}
& CISD     &   -0.6483671  &   -0.7593664  &   -0.8151218  &   -0.8395635  &    -0.87322(98) \\
& RCCSD(T) &   -0.7176900  &   -0.8435027  &   -0.9086709  &   -0.9373488  &     -0.9771(12) \\
& UCCSD(T) &   -0.7178303  &   -0.8436765  &   -0.9088664  &   -0.9375544  &     -0.9773(12) \\
& CCSDT(Q) &   -0.7175330  &   -0.8427576  &   -0.90787(*) &   -0.93653(*) &     -0.9763(12) \\
& ROHF     & -168.4166314  & -168.4169417  & -168.4169483  & -168.4169498  &  -168.416952(6) \\
\cline{7-7}
& Total    &               &               &               &               &   -169.3932(12) \\
\hline

\multirow{6}{*}{\bf{Cu}($^2$S)}
& CISD     &   -0.6998956  &   -0.8247607  &   -0.8882785  &   -0.9164908  &    -0.95506(89) \\
& RCCSD(T) &   -0.7738402  &   -0.9159891  &   -0.9900301  &   -1.0231374  &    -1.06846(99) \\
& UCCSD(T) &   -0.7739280  &   -0.9160963  &   -0.9901465  &   -1.0232562  &    -1.06859(99) \\
& CCSDT(Q) &   -0.7728824  &   -0.9141401  &   -0.98803(*) &   -1.02107(*) &     -1.0664(10) \\
& ROHF     & -195.3358266  & -195.3373699  & -195.3373989  & -195.3374008  &  -195.337402(3) \\
\cline{7-7}
& Total    &               &               &               &               &   -196.4038(10) \\
\hline

\multirow{6}{*}{\bf{Zn}($^1$S)}
& CISD     &   -0.7139072  &   -0.8425259  &   -0.9112300  &   -0.9413355  &      -0.9836(15) \\
& RCCSD(T) &   -0.7892886  &   -0.9339068  &   -1.0128710  &   -1.0475878  &      -1.0964(18) \\
& UCCSD(T) &   -0.7892884  &   -0.9339071  &   -1.0128710  &   -1.0475878  &      -1.0964(18) \\
& CCSDT(Q) &   -0.7881588  &   -0.93257(*) &   -1.01142(*) &   -1.04608(*) &      -1.0949(18) \\
& RHF      & -225.2750460  & -225.2750619  & -225.2750649  & -225.2750650  &  -225.2750654(6) \\
\cline{7-7}
& Total    &               &               &               &               &    -226.3699(18) \\

\hline\hline
\end{tabular}
\end{table*}

\begin{table*}[htbp!]
\setlength{\tabcolsep}{4pt} 
\centering
\small
\caption{
Atomic correlation and total energies [Ha] for the
3rd-row main group with ccECPs. Notation as in table \ref{1,ccECP,SA,en}.
aug-cc-pVnZ basis set (cc-pCVnZ for K, Ca).}
\label{3rd-main,ccECP,SA,en}
\begin{adjustbox}{width=0.85\textwidth,center}
\begin{tabular}{l|l|rrrrr|r}
\hline\hline
Atom & Method &          DZ &          TZ &          QZ &          5Z &       6Z     & CBS \\
\hline

\multirow{6}{*}{\bf{K}}
& CISD     &  -0.17702362 &  -0.25025818 &  -0.27680275 &  -0.28486555 &  -0.28829755 &   -0.29235(21) \\
& RCCSD(T) &  -0.18450036 &  -0.26859548 &  -0.29941325 &  -0.30862333 &  -0.31247228 &   -0.31694(21) \\
& UCCSD(T) &  -0.18450996 &  -0.26860642 &  -0.29942606 &  -0.30863645 &  -0.31248545 &   -0.31695(21) \\
& CCSDT(Q) &  -0.18489132 &  -0.26921595 &  -0.30016619 &  -0.30939(*) &  -0.31325(*) &   -0.31772(22) \\
& ROHF      & -27.93462232 & -27.93462232 & -27.93462245 & -27.93462245 & -27.93462530 &  -27.934626(4) \\
\cline{8-8}
& Total    &              &              &              &              &              &  -28.25235(22) \\
\hline

\multirow{6}{*}{\bf{Ca}}
& CISD     &  -0.22929612 &  -0.29416623 &  -0.32310943 &  -0.33133840 &  -0.33484191 &   -0.33854(38) \\
& RCCSD(T) &  -0.25051655 &  -0.32692010 &  -0.36059895 &  -0.37009280 &  -0.37412840 &   -0.37833(46) \\
& UCCSD(T) &  -0.25051659 &  -0.32692014 &  -0.36059897 &  -0.37009283 &  -0.37412851 &   -0.37833(46) \\
& CCSDT(Q) &  -0.25129831 &  -0.32773983 &  -0.36145323 &  -0.37096(*) &  -0.37501(*) &   -0.37924(46) \\
& RHF      & -36.34973416 & -36.34973414 & -36.34973418 & -36.34973433 & -36.34973523 &  -36.349736(2) \\
\cline{8-8}
& Total    &              &              &              &              &              &  -36.72897(46) \\
\hline

\multirow{6}{*}{\bf{Ga}}
& CISD     & -0.04721657 & -0.05204881 & -0.05334383 & -0.05373906 & -0.05388443 &  -0.054070(13) \\
& RCCSD(T) & -0.04802291 & -0.05328084 & -0.05468337 & -0.05511537 & -0.05527408 &  -0.055480(15) \\
& UCCSD(T) & -0.04803310 & -0.05333764 & -0.05474019 & -0.05517174 & -0.05533016 &  -0.055535(15) \\
& FCI      & -0.04832843 & -0.05367889 & -0.05504512 & -0.05545554 & -0.05560451 &  -0.055791(13) \\
& ROHF      & -1.98412429 & -1.98412425 & -1.98412427 & -1.98412431 & -1.98412435 &  -1.9841244(2) \\
\cline{8-8}
& Total    &             &             &             &             &             &  -2.039915(13) \\
\hline

\multirow{7}{*}{\bf{Ge}}
& CISD     & -0.06192765 & -0.07159052 & -0.07444651 & -0.07514909 & -0.07546205 &  -0.075697(82) \\
& RCCSD(T) & -0.06345073 & -0.07438139 & -0.07758408 & -0.07837227 & -0.07871955 &  -0.078981(88) \\
& UCCSD(T) & -0.06347933 & -0.07456887 & -0.07777267 & -0.07855857 & -0.07890468 &  -0.079163(89) \\
& CCSDT(Q) & -0.06386528 & -0.07518215 & -0.07832601 & -0.07906773 & -0.07939291 &  -0.079608(93) \\
& FCI      & -0.06387817 & -0.07519494 & -0.07833928 & -0.07908014 & -0.07940490 &  -0.079619(94) \\
& ROHF      & -3.66482420 & -3.66482408 & -3.66482412 & -3.66482426 & -3.66482436 &  -3.6648245(3) \\
\cline{8-8}
& Total    &             &             &             &             &             &  -3.744443(94) \\
\hline

\multirow{7}{*}{\bf{As}}
& CISD     & -0.07039641 & -0.08714420 & -0.09165502 & -0.09303262 & -0.09360150 &  -0.094288(16) \\
& RCCSD(T) & -0.07212776 & -0.09120004 & -0.09632986 & -0.09787671 & -0.09850951 &  -0.099262(18) \\
& UCCSD(T) & -0.07215698 & -0.09150397 & -0.09663415 & -0.09817850 & -0.09880856 &  -0.099556(17) \\
& CCSDT(Q) & -0.07246222 & -0.09216830 & -0.09725624 & -0.09874078 & -0.09933940 &  -0.100017(24) \\
& FCI      & -0.07248341 & -0.09218100 & -0.09727123 & -0.09875(*) & -0.09935(*) &  -0.100032(24) \\
& ROHF      & -6.06587734 & -6.06587663 & -6.06587765 & -6.06587778 & -6.06587820 &  -6.0658782(4) \\
\cline{8-8}
& Total    &             &             &             &             &             &  -6.165911(24) \\
\hline

\multirow{7}{*}{\bf{Se}}
& CISD     & -0.09384530 & -0.12606639 & -0.13651429 & -0.13953058 & -0.14076645 &  -0.142131(79) \\
& RCCSD(T) & -0.09800015 & -0.13409237 & -0.14577687 & -0.14911247 & -0.15046493 &  -0.151934(85) \\
& UCCSD(T) & -0.09806646 & -0.13429791 & -0.14603040 & -0.14936820 & -0.15071940 &  -0.152179(87) \\
& CCSDT(Q) & -0.09852407 & -0.13501159 & -0.14677118 & -0.15004249 & -0.15135002 &  -0.152709(93) \\
& FCI      & -0.09854509 & -0.13502983 & -0.14679328 & -0.15006(*) & -0.15137(*) &  -0.152731(93) \\
& ROHF      & -9.14758103 & -9.14761968 & -9.14762395 & -9.14763353 & -9.14763620 &    -9.14764(2) \\
\cline{8-8}
& Total    &             &             &             &             &             &  -9.300373(95) \\
\hline

\multirow{7}{*}{\bf{Br}}
& CISD     &  -0.11114459 &  -0.15568329 &  -0.17487123 &  -0.17998183 &  -0.18203497 &   -0.18397(23) \\
& RCCSD(T) &  -0.11646859 &  -0.16645788 &  -0.18792038 &  -0.19358220 &  -0.19582847 &   -0.19791(25) \\
& UCCSD(T) &  -0.11651773 &  -0.16655839 &  -0.18807691 &  -0.19374303 &  -0.19598914 &   -0.19806(25) \\
& CCSDT(Q) &  -0.11694920 &  -0.16719402 &  -0.18879592 &  -0.19441073 &  -0.19666(*) &   -0.19867(30) \\
& FCI      &  -0.11696191 &  -0.16720976 &              &              &              &                \\
& ROHF      & -13.11960933 & -13.11961798 & -13.11961590 & -13.11961959 & -13.11962121 &  -13.119624(7) \\
\cline{8-8}
& Total    &              &              &              &              &              &  -13.31830(30) \\
\hline

\multirow{7}{*}{\bf{Kr}}
& CISD     &  -0.13816713 &  -0.18591690 &  -0.21246190 &  -0.21982064 &  -0.22280036 &   -0.22586(26) \\
& RCCSD(T) &  -0.14505536 &  -0.19969369 &  -0.22916475 &  -0.23733449 &  -0.24060491 &   -0.24398(25) \\
& UCCSD(T) &  -0.14505533 &  -0.19969370 &  -0.22916475 &  -0.23733449 &  -0.24060491 &   -0.24398(25) \\
& CCSDT(Q) &  -0.14545281 &  -0.20023839 &  -0.22975824 &  -0.23790612 &  -0.24118(*) &   -0.24453(27) \\
& FCI      &  -0.14546095 &              &              &              &              &                \\
& RHF      & -18.22805984 & -18.22806066 & -18.22806022 & -18.22806208 & -18.22806315 &  -18.228065(5) \\
\cline{8-8}
& Total    &              &              &              &              &              &  -18.47259(27) \\

\hline\hline
\end{tabular}
\end{adjustbox}
\end{table*}

\begin{table*}[htbp!]
\setlength{\tabcolsep}{4pt} 
\centering
\small
\caption{Atomic correlation and total energies [Ha] of selected $3d$ elements with STU[Ne] ECPs. Notation as in table \ref{1,ccECP,SA,en}.  cc-pCVnZ basis set.}
\label{3,STU,SA,en}
\begin{tabular}{l|l|rrrr|r}
\hline\hline
Atom & Method &    DZ &      TZ &          QZ &          5Z &             CBS \\
\hline

\multirow{6}{*}{\bf{Fe}($^5$D)}
& CISD     &   -0.5119665  &   -0.5906234  &   -0.6276620  &   -0.6435827  &    -0.66542(72) \\
& RCCSD(T) &   -0.5667378  &   -0.6561334  &   -0.6986089  &   -0.7166818  &     -0.7418(10) \\
& UCCSD(T) &   -0.5672400  &   -0.6566930  &   -0.6992014  &   -0.7172884  &     -0.7424(10) \\
& CCSDT(Q) &   -0.5679571  &   -0.6572278  &   -0.6998(*)  &   -0.7179(*)  &     -0.7430(10) \\
& ROHF     & -123.1145055  & -123.1145060  & -123.1145058  & -123.1145106  &  -123.114511(6) \\
\cline{7-7}
& Total    &               &               &               &               &   -123.8575(10) \\
\hline

\multirow{6}{*}{\bf{Co}($^4$F)}
& CISD     &   -0.5587796  &   -0.6510040  &   -0.6953698  &   -0.7146899  &    -0.74108(74) \\
& RCCSD(T) &   -0.6177796  &   -0.7223349  &   -0.7733089  &   -0.7953192  &     -0.8257(10) \\
& UCCSD(T) &   -0.6181501  &   -0.7227785  &   -0.7737868  &   -0.7958146  &     -0.8263(10) \\
& CCSDT(Q) &   -0.6185380  &   -0.7229430  &   -0.7740(*)  &   -0.7960(*)  &     -0.8265(11) \\
& ROHF     & -144.9403448  & -144.9403454  & -144.9403450  & -144.9403522  &  -144.940354(7) \\
\cline{7-7}
& Total    &               &               &               &               &   -145.7668(11) \\
\hline

\multirow{6}{*}{\bf{Ni}($^3$F)}
& CISD     &   -0.6075365  &   -0.7127520  &   -0.7643948  &   -0.7870949  &    -0.81806(77) \\
& RCCSD(T) &   -0.6716907  &   -0.7904285  &   -0.8497909  &   -0.8757266  &     -0.9115(11) \\
& UCCSD(T) &   -0.6719212  &   -0.7907215  &   -0.8501096  &   -0.8760587  &     -0.9118(11) \\
& CCSDT(Q) &   -0.6719084  &   -0.7904379  &   -0.8498(*)  &   -0.8757(*)  &     -0.9115(11) \\
& ROHF     & -169.9419868  & -169.9419873  & -169.9419869  & -169.9419936  &  -169.941995(8) \\
\cline{7-7}
& Total    &               &               &               &               &   -170.8535(11) \\
\hline\hline
\end{tabular}
\end{table*}

\section{Correlation energies from selected-CI method}

Here we expand our study by employing another accurate method
that is proving very valuable for 
generating trial functions for QMC
calculations.
To this end, we carried out selected-CI calculations which utilize CIPSI (Configuration Interaction using a Perturbative Selection made Iteratively) algorithm\cite{CIPSI} for some selected cases.
Table \ref{1,ccECP,CIPSI} shows selected ccECP cases and their correlation energies with an increasing number of determinants using this method.
Similarly, we also carried out selected-CI calculations for ccECP[Ne] TMs in table \ref{3,ccECP,CIPSI} which provides variational energies as well as energies with second-order perturbation corrections (E+PT2) from multi-determinant expansions. 
Corresponding DMC calculations using trial functions with different numbers of determinants from applying truncation thresholds  are given later, in table \ref{3,ccECP,CI-DMC}.

\begin{table}[htbp!]
\setlength{\tabcolsep}{4pt} 
\centering
\caption{Selected ccECP[He] correlation energies [Ha] from selected-CI (CIPSI) method with cc-pVnZ basis sets.}
\label{1,ccECP,CIPSI}
\begin{adjustbox}{width=\columnwidth,center}
\begin{tabular}{c|rrcc}
\hline\hline

Atom & Basis      &     \# Dets &  E(Variational) & "Exact" \\
\hline
     &    QZ      &       1714 &   -0.07504964 & \multirow{6}{*}{-0.07608(13)} \\
     &            &       3214 &   -0.07514184 \\
  B  &            &       5628 &   -0.07515945 \\
     &    5Z      &       2578 &   -0.07539506 \\
     &            &       5138 &   -0.07554463 \\
     &            &       9696 &   -0.07558846 \\
\hline
     &    QZ      &       6463 &   -0.10074050 & \multirow{6}{*}{-0.10328(31)} \\
     &            &      13636 &   -0.10092427 \\
  C  &            &      27139 &   -0.10096552 \\
     &    5Z      &      16028 &   -0.10184902 \\
     &            &      34543 &   -0.10198856 \\
     &            &      70384 &   -0.10202771 \\
\hline
     &    QZ      &      64881 &   -0.12758307 & \multirow{6}{*}{-0.132131(85)} \\
     &            &     148521 &   -0.12780090 \\
  N  &            &     308137 &   -0.12785924 \\
     &    5Z      &     131957 &   -0.12959808 \\
     &            &     312200 &   -0.12979673 \\
     &            &     688544 &   -0.12986038 \\
\hline
     &    QZ      &      74704 &   -0.18411219 & \multirow{6}{*}{-0.19760(20)} \\
     &            &     200813 &   -0.18456574 \\
  O  &            &     455639 &   -0.18461709 \\
     &    5Z      &     300936 &   -0.18910706 \\
     &            &     809834 &   -0.18934903 \\
     &            &    1864624 &   -0.18937439 \\

\hline\hline
\end{tabular}
\end{adjustbox}
\end{table}

\begin{table}[!htbp]
\setlength{\tabcolsep}{4pt} 
\centering
\small
\caption{Transition metal atoms ccECP[Ne]  total energies [Ha] using CIPSI method. cc-pCVQZ basis set without $h$ functions was used. Determinants are given in millions (M). 
Natural orbitals were obtained from a wave function with approximately 0.15M determinants.}
\label{3,ccECP,CIPSI}
\begin{adjustbox}{width=\columnwidth,center}
\begin{tabular}{l|cccc}
\hline\hline
Atom & \# Dets(M) &  E(Variational) & E+PT2 & "Exact"\\
\hline

\multirow{3}{*}{\bf{Sc}($^2$D)}
&   0.28   & -46.52014938 & -46.53496469 \\
&   1.03   & -46.52603096 & -46.53483770 & -46.55704(81) \\
&   4.11   & -46.52986709 & -46.53475117 \\
\hline

\multirow{3}{*}{\bf{Ti}($^3$F)}
&   0.12   & -58.03295957 & -58.06347742 \\
&   0.51   & -58.04482977 & -58.06336651 & -58.09263(76) \\
&   2.38   & -58.05267212 & -58.06319320 \\
\hline

\multirow{3}{*}{\bf{V}($^4$F)}
&   0.09   & -71.35079260 &  -71.39666941 \\
&   0.38   & -71.36911296 &  -71.39863887 & -71.44178(59) \\
&   1.96   & -71.38282921 &  -71.40112340 \\
\hline

\multirow{3}{*}{\bf{Cr}($^7$S)}
&   0.11   & -86.54572850 & -86.59312840 \\
&   0.55   & -86.56552838 & -86.59188852 & -86.64109(33) \\
&   3.59   & -86.57703357 & -86.59154867 \\
\hline

\multirow{3}{*}{\bf{Mn}($^6$S)}
&   0.15   & -103.75114683 & -103.84014855 \\
&   0.64   & -103.78416151 & -103.83722059 & -103.8919(10) \\
&   3.15   & -103.80379565 & -103.83794337 \\
\hline

\multirow{3}{*}{\bf{Fe}($^5$D)}
&   0.15   & -123.25398200 & -123.31791337 \\
&   0.80   & -123.27957505 & -123.31957722 & -123.38804(93) \\
&   5.52   & -123.29635557 & -123.31979931 \\
\hline

\multirow{3}{*}{\bf{Co}($^4$F)}
&  0.18   & -145.00085764 & -145.06979414 \\
&  0.91   & -145.02672886 & -145.07174210 & -145.1541(10) \\
&  6.11   & -145.04447415 & -145.07223050 \\
\hline

\multirow{3}{*}{\bf{Ni}($^3$F)}
&  0.13   & -169.20259118 & -169.28369714 \\
&  0.54   & -169.23066204 & -169.28809650 & -169.3912(12) \\
&  2.88   & -169.25254776 & -169.29256526 \\
\hline

\multirow{3}{*}{\bf{Cu}($^2$S)}
&  0.21   & -196.21649194 & -196.28741767 \\
&  1.00   & -196.23676320 & -196.28659911 & -196.4038(10) \\
&  5.36   & -196.25356865 & -196.28606411 \\
\hline

\multirow{3}{*}{\bf{Zn}($^1$S)}
& 0.21  & -226.17497021  & -226.24479408 \\
& 0.66  & -226.19040192  & -226.24552355 & -226.3699(18) \\
& 2.28  & -226.20447811  & -226.24564665 \\
\hline

\hline\hline
\end{tabular}
\end{adjustbox}
\end{table}

\section{Fixed-node DMC calculations and  corresponding bias}\label{FNDMC}

It is well-known that the systematic errors in DMC ECP/pseudopotential calculations correspond to the fixed-node (FN) and localization errors \cite{Mitas1991}.
We note the obvious fact that the localization error is not present for ECPs with only local operators such as H and He.
Moreover, systems such as Be with He core (Be[He]) or Mg[Ne] have no FN errors because the $^1S[2s^2]$ state with $2s$ pseudo-orbitals is nodeless.
Therefore, the errors in these cases correspond to the localization error only. 
For instance, for Be[He], we find the localization error to be 1.3(2) mHa for HF nodes and 1.7(3) mHa for PBE nodes.

\begin{figure}[htbp!]
\caption{
Comparison of total energies using different methods and basis sets (cc-pVnZ) for F atom with ccECP[He].
Lower part of the figure expands the region
around the SJ DMC and exact values. We include CIPSI variational, CIPSI E+PT2 and DMC. 
HF reference was used for the CIPSI expansions.
Fixed-node DMC values with CIPSI multi-reference wave functions based on nZ bases do not include the Jastrow factor and result from
extrapolations with $\tau=(0.01, 0.005, 0.0025)$ Ha$^{-1}$ time steps.
SJ DMC represents DMC energy of the single-reference trial function with the Jastrow factor included. See text for further details on the methods involved.
}
\includegraphics[width=\columnwidth]{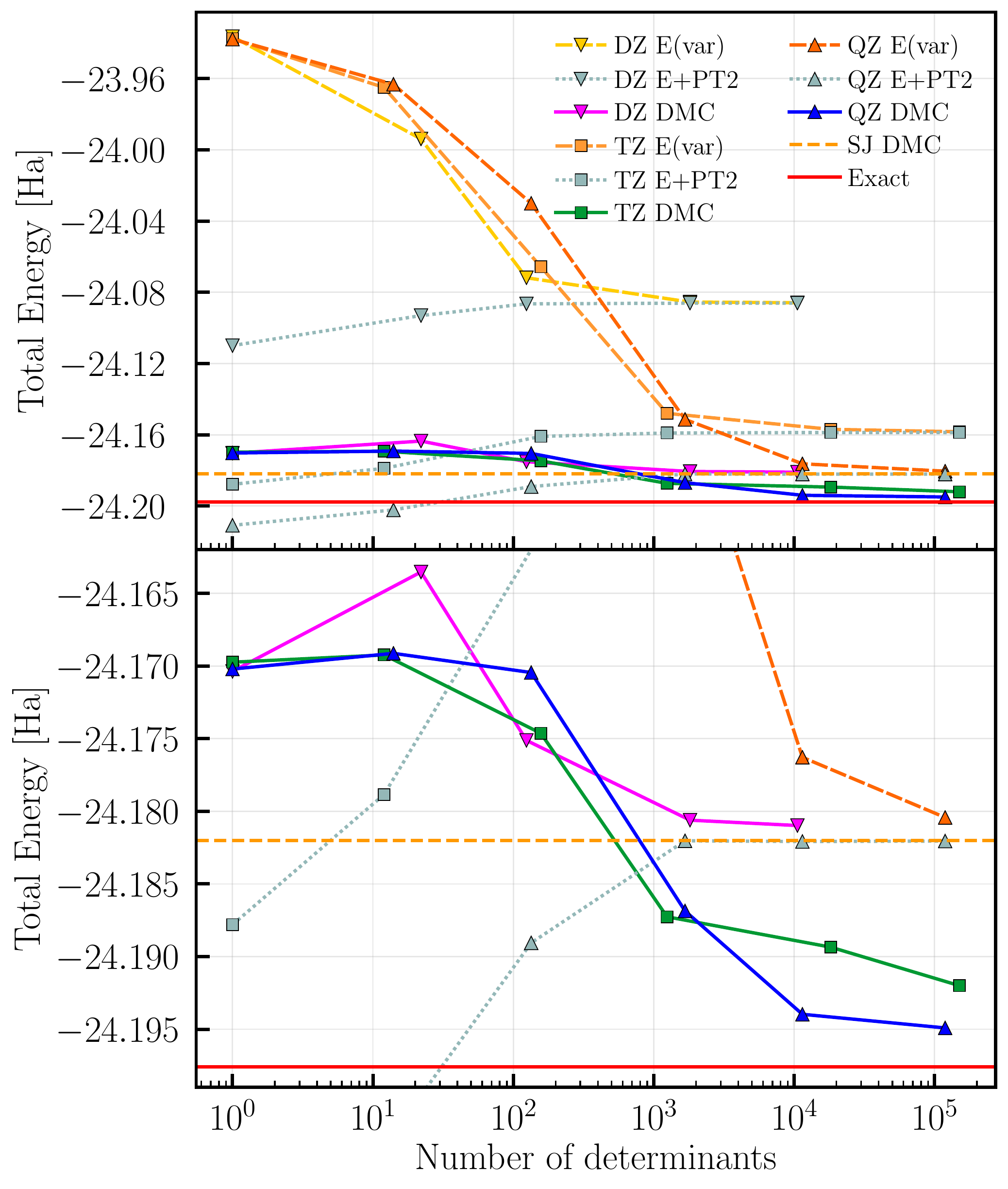}
\label{fig:comp_methods}
\centering
\end{figure}

We can also observe that the quality of the basis set used in DMC calculations does not significantly affect the energy of single reference atomic systems. 
This is because for atoms, the nodes of the HF or single determinant wave functions are essentially fully captured by accurately contracted basis set at the DZ level.
We illustrate this fact in table \ref{tab:DMC_vs_basis} for selected atoms from different rows, with and without T-moves, where we see that the energies agree within the error bars.
We also want to note that this does not apply to molecular systems where TZ level is needed and importance of higher angular momenta orbitals is more pronounced.
Similarly, on an example of the fluorine atom, we illustrate the dependence of atomic DMC energies on the basis set when the corresponding multi-determinant CIPSI expansions are employed as shown in figure \ref{fig:comp_methods}.

\begin{table}[!htbp]
\setlength{\tabcolsep}{4pt} 
\centering
\caption{Selected ccECP pseudopotential single-reference fixed-node DMC energies [Ha] probing for impact of basis and T-moves. Note that the energies of an atom at different basis sets agree within the error bars. Extrapolated with $\tau=(0.02, 0.01, 0.005, 0.0025)$ Ha$^{-1}$ time steps.}
\label{tab:DMC_vs_basis}
\begin{adjustbox}{width=\columnwidth,center}
\begin{tabular}{cc|cccc}
\hline\hline
Atom & T-Moves & DZ & TZ & QZ & 5Z\\
\hline
O   & Yes & -15.8697(2)   & -15.8696(2)  & -15.8693(2)  & -15.8693(2) \\
O   & No  & -15.8720(3)   & -15.8719(4)  & -15.8727(3)  & -15.8728(4)  \\
Si  & Yes & -3.7601(1)    & -3.7601(1)   & -3.7602(1)   & 3.7600(1) \\
Si  & No  & -3.7602(1)    & -3.7604(1)   & -3.7605(1)   & -3.7601(1)   \\
Co  & Yes & -145.0711(3)  & -145.0708(3) & -145.0711(3) & -145.0706(3) \\
Co  & No  & -145.0755(3)  & -145.0751(3) & -145.0753(4) & -145.0748(4)  \\
\hline
\hline
\end{tabular}
\end{adjustbox}
\end{table}

Most accurate total energies for ccECP[He] 1st-row elements along with fixed-node DMC energies with single-reference HF trial wave functions are given in table \ref{tab:1st_ultimate}.
Similarly, these data for 2nd row are given tables \ref{tab:2nd_ultimate} and \ref{tab:2nd-He_ultimate} for [Ne] and [He] cores respectively.
Note that the locality error in Mg[Ne] is an order of magnitude smaller than Be[He] since there is only very mild effect from $s^2\to p^2$ configuration mixing in the 2nd row. 

The increase in both absolute and relative values of the fixed-node bias for [He] core is driven by the dominant energy scale of the semicore subshell $2s^22p^6$. It must go up due to the corresponding increase in the electronic density which pushes the fixed-node error to larger values as explained elsewhere \cite{Rasch2012, Rasch2014}. 
One can expect that in energy differences this contribution will mostly cancel out.  Note that the valence component of the correlation energy remains essentially the same while relatively getting smaller and most probably less affected by the core effects. This has been seen also in comparisons of QMC all-electron and ECP calculations that show very similar
trends in differences\cite{Needs_2010}. Obviously, further explorations in this direction will be very useful and should better quantify this particular aspect of error cancellations.

\begin{table}[!htbp]\centering
\setlength{\tabcolsep}{4pt} 
\caption{
Most accurate total energies for ccECP[He] 1st-row elements along with fixed-node DMC energies with single-reference HF trial wave functions   
based on data from $\tau=(0.02, 0.01, 0.005, 0.0025)$ Ha$^{-1}$ time step extrapolation.
Here, $\eta=(100 \epsilon)/|E_{corr}|$ where $\epsilon$ represents the total DMC error, namely, combined fixed-node and localization biases.}

\label{tab:1st_ultimate}
\begin{adjustbox}{width=\columnwidth,center}
\begin{tabular}{c|cccc}
\hline\hline
Atom & "Exact" (Ha)   & DMC/HF(Ha)    & $\epsilon$(mHa) &  $\eta$ \\
\hline
H  &  -0.49999965(1)  &               &           &           \\
He &  -2.903745(23)   & -2.903728(1)  &  0.02(2)  &  0.04(5)  \\
Li &  -0.19685279(1)  &               &           &           \\
Be &  -1.01023896(12) & -1.0089(2)    &  1.3(2)   &  2.8(4)   \\
B  &  -2.61531(13)    & -2.6061(1)    &  9.2(2)   &  12.1(2)  \\
C  &  -5.41753(31)    & -5.4052(2)    &  12.3(4)  &  11.9(4)  \\
N  &  -9.765999(85)   & -9.7551(2)    &  10.9(2)  &   8.2(2)  \\
O  &  -15.88439(20)   & -15.8696(2)   &  14.8(3)  &   7.5(1)  \\
F  &  -24.1976(11)    & -24.1820(2)   &  16(1)    &   6.0(4)  \\
Ne &  -35.03250(96)   & -35.0210(2)   &  11(1)    &   3.6(3)  \\

\hline
\hline
\end{tabular}
\end{adjustbox}
\end{table}

\begin{table}[!htbp]
\centering
\caption{
Most accurate total energies for ccECP[Ne] 2nd-row elements along with fixed-node DMC energies with single-reference HF trial wave functions.   
Notations as in table \ref{tab:1st_ultimate}.}
\label{tab:2nd_ultimate}
\begin{adjustbox}{width=\columnwidth,center}
\begin{tabular}{c|cccc}
\hline\hline
Atom & "Exact"(Ha)      & DMC/HF(Ha)  & $\epsilon$(mHa) & $\eta$ \\
\hline
Na  & -0.186206(1)    &             &         &          \\
Mg  & -0.823473(30)   & -0.8234(1)  &  0.1(1) &  0.2(3)  \\
Al  & -1.9375235(79)  & -1.9366(1)  &  0.9(1) &  1.5(2)  \\
Si  & -3.762073(57)   & -3.7601(1)  &  2.0(1) &  2.2(1)  \\
P   & -6.459803(74)   & -6.4577(1)  &  2.1(1) &  1.8(1)  \\
S   & -10.09742(13)   & -10.0920(2) &  5.4(2) &  3.0(1)  \\
Cl  & -14.92670(18)   & -14.9188(2) &  7.9(3) &  3.3(1)  \\
Ar  & -21.07170(24)   & -21.0632(2) &  8.5(3) &  2.9(1)  \\
\hline
\hline
\end{tabular}
\end{adjustbox}
\end{table}

\begin{table}[!htbp]
\centering
\caption{
Most accurate total energies for ccECP[He] 2nd-row elements along with fixed-node DMC energies with single-reference HF trial wave functions. 
Notations as in table \ref{tab:1st_ultimate}.}

\label{tab:2nd-He_ultimate}
\begin{adjustbox}{width=\columnwidth,center}
\begin{tabular}{c|cccc}
\hline\hline
Atom & "Exact"(Ha)      & DMC/HF(Ha)  & $\epsilon$(mHa) & $\eta$ \\
\hline
Na  & -47.680977(64)   & -47.6668(2)  & 14.2(2) &  4.38(6) \\
Mg  & -63.28914(13)    & -63.2713(4)  & 17.8(4) &   4.9(1) \\
Al  & -81.38579(37)    & -81.3621(5)  & 23.7(6) &   6.0(2) \\
Si  & -102.0512597(35) & -102.0230(4) & 28.3(4) &  6.65(9) \\
P   & -125.717593(74)  & -125.6836(5) & 34.0(5) &   7.4(1) \\
S   & -152.439331(87)  & -152.3982(4) & 41.1(4) &  7.87(8) \\
Cl  & -182.19712(21)   & -182.1476(5) & 49.5(5) &  8.48(9) \\
Ar  & -215.53238(18)   & -215.4738(5) & 58.6(5) &  9.15(8) \\
\hline
\hline
\end{tabular}
\end{adjustbox}
\end{table}

The fixed-node DMC atomic energies for the K-Zn ccECP[Ne] and Ga-Kr ccECP[[Ar]3$d^{10}$] pseudoatoms are given in tables \ref{tab:3rd_ultimate} and \ref{tab:4p_ultimate}. 
Note that for Ni, we provide two energies corresponding to different states of [Ar]$3d^84s^2$ ($^3F$) and [Ar]$3d^94s^1$ ($^3D$).
Ni atom is a special case since it displays the so-called $s \leftrightarrow d$ instability and the mentioned states are energetically very close according to the experiments \cite{Ni-instability, NIST_ASD}. 
We find that these energies are degenerate within the obtained errors, although we observe that the $^3D$ state has somewhat lower fixed-node bias.

\begin{table}[!htbp]
    \centering
    \caption{
    Most accurate total energies for ccECP[Ne] K-Zn elements along with fixed-node DMC energies with single-reference HF trial wave functions.  
    Notations as in table \ref{tab:1st_ultimate}.}
    
    \label{tab:3rd_ultimate}
    \begin{adjustbox}{width=\columnwidth,center}
    \begin{tabular}{cl|cccc}
        \hline\hline
        Atom & State & "Exact"(Ha) & DMC/HF(Ha) & $\epsilon$(mHa) & $\eta$ \\
        \hline
        K   & ($^2$S) & -28.25243(25)   & -28.2394(2)   &  13.0(3)  &   4.1(1) \\
        Ca  & ($^1$S) & -36.72897(46)   & -36.7055(2)   &  23.5(5)  &   6.2(1) \\
        Sc  & ($^2$D) & -46.55704(81)   & -46.5202(4)   &  36.8(9)  &   8.4(2) \\
        Ti & ($^3$F)  & -58.09263(76)   & -58.0458(2)   & 46.8(8)   &   9.6(2) \\
        V  & ($^4$F)  & -71.44178(59)   & -71.3829(2)   & 58.9(6)   &  10.8(1) \\
        Cr  & ($^7$S) & -86.64109(33)   & -86.5876(2)   &  53.5(4)  &  9.03(7) \\
        Mn  & ($^6$S) & -103.8919(10)   & -103.8260(3)  &  66(1)    &  10.2(2) \\
        Fe  & ($^5$D) & -123.38804(93)  & -123.3100(3)  &  78(1)    &  10.5(1) \\
        Co & ($^4$F)  & -145.1541(10)   & -145.0709(3)  &   83(1)   &  10.1(1) \\
        Ni & ($^3$F)  & -169.3912(12)   & -169.2973(6)  &   94(1)   &  10.3(1) \\
        Ni  & ($^3$D) & -169.3932(12)   & -169.3056(6)  &  88(1)    &   9.0(1) \\
        Cu  & ($^2$S) & -196.4038(10)   & -196.3178(3)  &  86(1)    &   8.1(1) \\
        Zn  & ($^1$S) & -226.3699(18)   & -226.2775(4)  &  92(2)    &   8.4(2) \\

        \hline
        \hline
    \end{tabular}
    \end{adjustbox}
\end{table}

\begin{table}[!htbp]
    \centering
    \caption{
    Most accurate total energies for ccECP[[Ar]3$d^{10}$] Ga-Kr elements along with fixed-node DMC energies with single-reference HF trial wave functions.  
    Notations as in table \ref{tab:1st_ultimate}.}
    
    \label{tab:4p_ultimate}
    \begin{adjustbox}{width=\columnwidth,center}
    \begin{tabular}{c|cccc}
        \hline\hline
        Atom & "Exact"(Ha) & DMC/HF(Ha) & $\epsilon$(mHa) & $\eta$ \\
        \hline
        Ga & -2.039915(13) &  -2.0392(2)  &   0.7(2)  &   1.3(4) \\
        Ge & -3.744443(94) &  -3.7429(3)  &   1.5(3)  &   1.9(4) \\
        As & -6.165911(24) &  -6.1637(2)  &   2.2(2)  &   2.2(2) \\
        Se & -9.300373(95) &  -9.2966(1)  &   3.8(1)  &  2.47(9) \\
        Br & -13.31830(30) &  -13.3138(1) &   4.5(3)  &   2.3(2) \\
        Kr & -18.47259(27) &  -18.4680(1) &   4.6(3)  &   1.9(1) \\
        \hline
        \hline
    \end{tabular}
    \end{adjustbox}
\end{table}


Table \ref{1,ccECP,CI-DMC} shows selected DMC energies for various single-reference nodes based on orbitals from HF, PBE, and PBE0 calculations.
It also tabulates DMC energies using multi-determinant expansions truncated at $10^{-8}$ threshold\footnote{See reference \cite{QPackage} for details of this truncation threshold.} with (sCI(8)J) and without Jastrow factors (sCI(8)).
We see a good agreement between "sCI(8)J" and previously tabulated (table \ref{1,ccECP,SA,en}) correlation energies, especially for B and C atoms where the energies are essentially the same. 

\begin{table}[!htbp]
\centering
\caption{DMC total energies [Ha] of selected ccECP[He] 1st-row elements with single-reference 
and multi-reference trial wave functions.}
\label{1,ccECP,CI-DMC}
\begin{adjustbox}{width=\columnwidth,center}
\begin{tabular}{l|llll}
\hline\hline
Atom        &            B &            C &            O &            N \\
\hline
DMC/HF          &   -2.6056(2) &   -5.4050(2) &  -15.8697(2) &   -9.7551(2) \\
DMC/PBE         &   -2.6053(2) &   -5.4055(2) &  -15.8663(3) &   -9.7545(2) \\
DMC/PBE0        &   -2.6053(2) &   -5.4044(2) &  -15.8697(3) &   -9.7545(2) \\
DMC/sCI(8)      &  -2.61524(6) &  -5.41728(7) &  -15.8822(1) &  -9.76545(7) \\
DMC/sCI(8)J     &  -2.61535(3) &  -5.41734(4) &  -15.8830(1) &  -9.76553(8) \\
"Exact"         & -2.61531(13) & -5.41753(31) & -15.88439(20)& -9.765999(85) \\
\hline
\hline
\end{tabular}
\end{adjustbox}
\end{table}

A DMC calculation of Fe atom is also shown for STU ECP in table \ref{STU_Fe_DMC}.
We find that for Fe atom, the sum of fixed-node and localization errors for ccECP and STU ECPs are almost the same within statistical deviations (78(1) mHa and 80(1) mHa, respectively).

\begin{table}
\centering
\caption{Fixed-node DMC energies [Ha] of the Fe atom with STU[Ne] ECP. $\tau=(0.01, 0.005, 0.0025, 0.001)$ Ha$^{-1}$ time step extrapolation was used. Corresponding VMC energy is 123.7270(4).}
\label{STU_Fe_DMC}
\begin{adjustbox}{width=0.7\columnwidth,center}
\begin{tabular}{l|ll}
\hline\hline
Timestep & T-moves & No T-moves \\
\hline
0.02     & -123.7984(5)   & -123.7981(6)  \\
0.01     & -123.7883(5)   & -123.7885(5)  \\
0.005    & -123.7838(4)   & -123.7848(4)  \\
0.0025   & -123.7807(4)   & -123.7846(4)  \\
0.001    & -123.7776(6)   & -123.7837(6)  \\
\hline 
Extrap.  & -123.7777(8)   & -123.7830(6) \\
\hline
\hline
\end{tabular}
\end{adjustbox}
\end{table}

\begin{figure*}[!htbp]
\centering
\begin{subfigure}{0.5\textwidth}
\includegraphics[width=\textwidth]{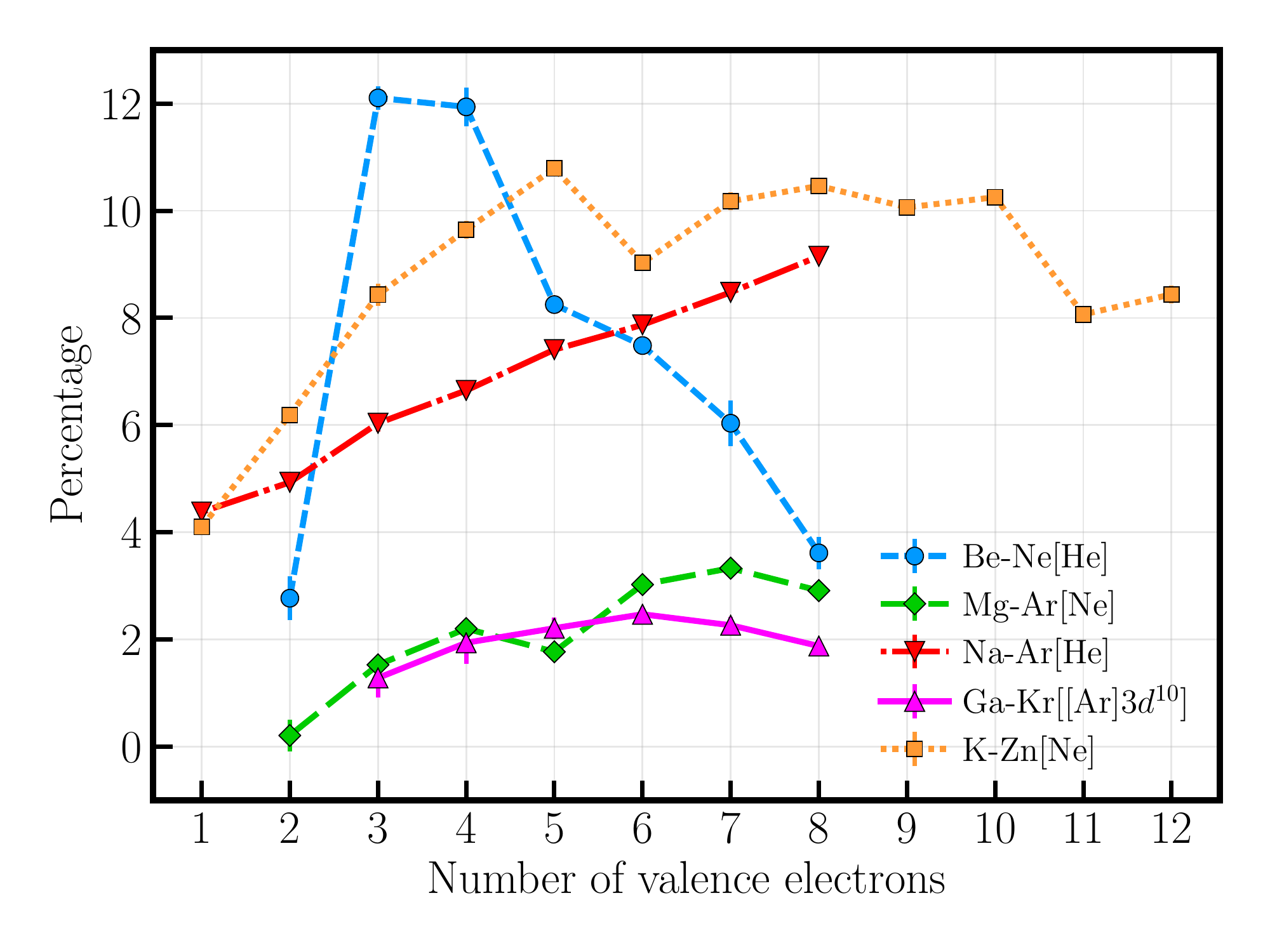}
\caption{}
\label{fig:FN_a}
\end{subfigure}%
\begin{subfigure}{0.5\textwidth}
\includegraphics[width=\textwidth]{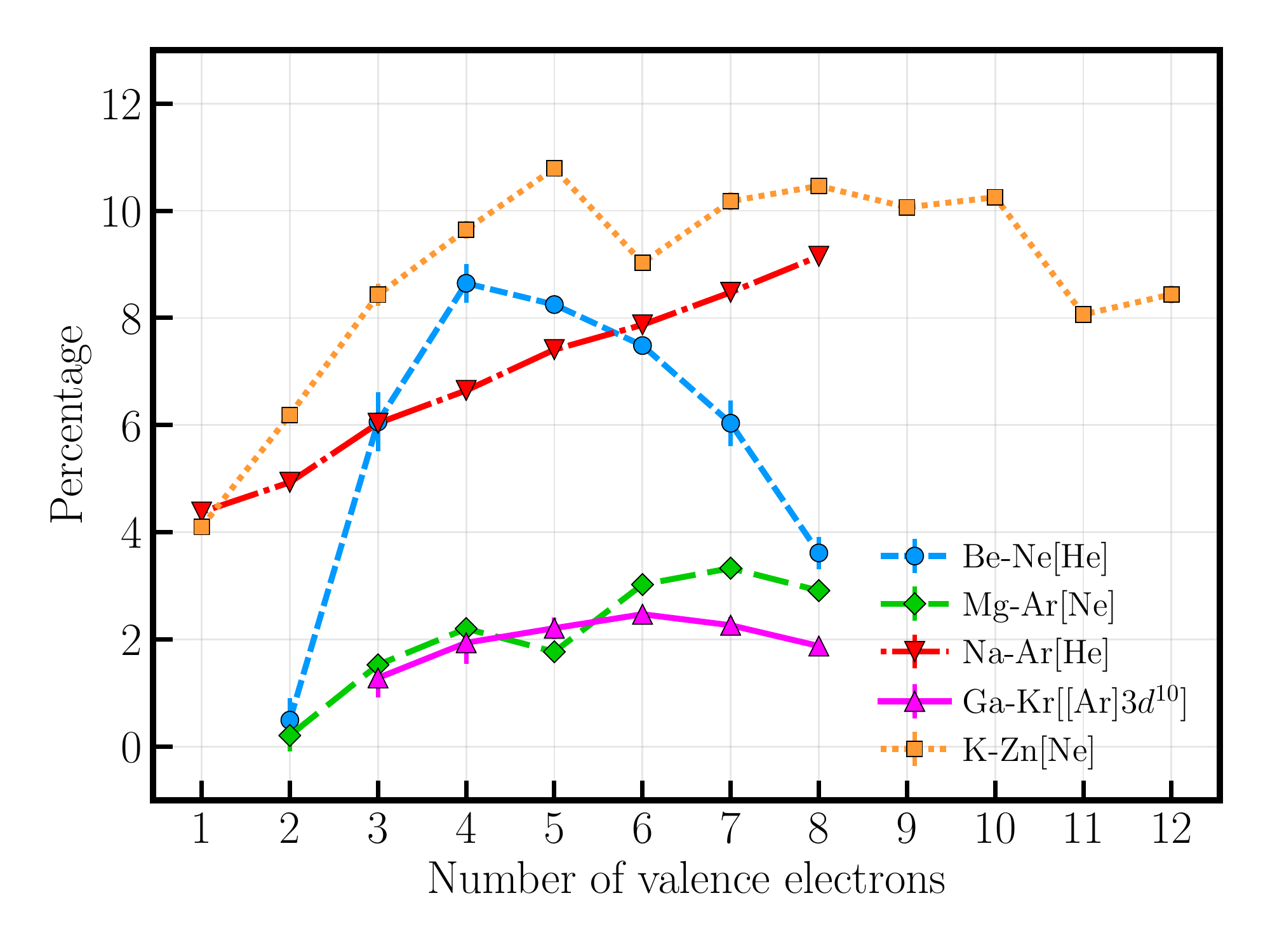}
\caption{}
\label{fig:FN_b}
\end{subfigure}
\caption{Fixed-node DMC biases as a percentage of the correlation energy for ccECP pseudoatoms: $100\epsilon/|E_{corr}|$. T-moves were used in calculations.
Part (a) shows results for single-reference trial functions. 
In part (b), Be, B, and C are calculated  with two-reference trial functions to account for the significant $2s-2p$ near-degeneracy.
See the text for further details.
}
\label{fig:FN}
\end{figure*}

A summary of systematic errors of the fixed-node DMC method is plotted in Fig. \ref{fig:FN} for all elements and core sizes considered in this work. 
In Fig. \ref{fig:FN_a}, we plotted the biases from the single reference HF trial wave functions. Perhaps surprisingly,
some of the largest percentage errors appear for B and C atoms. However, biases for these two elements and also for Be[He] are significantly
enhanced by the near-degeneracy effect between
$2s,2p$ levels. Such an effect is absent 
in the rest of the plotted elements (we note that although seemingly the same argument applies 
to the second and third row levels such as $3s,3p$, etc,
the corresponding values there are very marginal, below 
$\approx$ 0.5 mHa).
Therefore for Be, B, and C we have carried out calculations with two-configuration trial 
functions \cite{Umrigar,Rasch2012} and clearly
the errors decrease by several percents as shown in Fig. \ref{fig:FN_b}. 
For Be, the remaining small error of the order 
of 1\% or so,  is generated solely by the localization approximation as mentioned also above.  
It is interesting that the largest relative error appears for V atom. This reflects the large number of configurations 
that mix with the ground state due to the partial occupation of the $d$-shell as well as a significant contribution from correlations 
of $3d$ and semicore $3s,3p$ subshells. 
Note that for Cr that is the next element to V, the error rather abruptly decreases by about one fifth when compared with the V atom.  
This can be understood by much smaller 
number of excitations that mix with the high symmetry ground state due 
to the half-filled $d-$shell, $S$ total spatial angular
momentum and correspondingly
higher accuracy of the HF wave function due to 
high spin (septet) of the ground state. Also quite 
remarkably, the percentage error does not grow with the increasing
double occupancy of the $d-$shell where the main contribution to the correlation energy is from unlike spin pairs.  Since this 
is the domain of the Hubbard $U$ correlations we
see that these many-body effects are captured very well and quite consistently by the FN-DMC method as has been 
known for quite some time \cite{Wagner_2003_TM}.

The results are also consistent with previous
study on the extent and origins of fixed-node errors in first- (C, N, O)  vs second-row (Si, P, S) 
systems \cite{Rasch2014}. The graph shows that the fixed-node errors for these elements are about 2-3 \% and the lowest bias of about 1.5\% is found for the P atom.

In order to probe for improvement of 
the nodal surfaces especially for 
significant correlation effects in transition elements, we have carried out
fixed-node DMC with trial functions that 
included a large number of determinants
from selected-CI wave functions, see table \ref{3,ccECP,CI-DMC}.
 In this case, the Jastrow factor has not been used since the trial functions were very close to the exact ones. We see a significant improvement that decreases the fixed-node errors to about 4\% for V and Fe, and about 5\% for heavier transition atoms. 
We conjecture that the remaining bias is 
mainly in the semi-core $3s,3p$ channels
due to very large electron density in that region \cite{Rasch2014}.

\begin{table}[htbp!]
\setlength{\tabcolsep}{4pt} 
\centering
\small
\caption{Transition metal atoms ccECP[Ne] total energies [Ha] from the fixed-node DMC with multi-reference wave functions. cc-pCVQZ basis set without $h$ functions was used. The CI expansions were truncated at the indicated treshold with resulting number of determinants given in millions (M). Natural orbitals were obtained from a wave function with approximately 0.15M determinants. No Jastrow factor was used in the trial wave function. Timestep extrapolated with $\tau$=0.02,0.01,0.005 Ha$^{-1}$.}
\label{3,ccECP,CI-DMC}
\begin{adjustbox}{width=\columnwidth,center}
\begin{tabular}{l|cccc}
\hline\hline
Atom & Trunc. & \# Dets(M) & DMC & "Exact" \\
\hline

\multirow{2}{*}{\bf{Sc}($^2$D)}
&   1e-7   & 0.818 & -46.551(1) & \multirow{2}{*}{-46.55704(81)} \\
&   1e-8   & 1.834 & -46.550(1) \\
\hline

\multirow{2}{*}{\bf{Ti}($^3$F)}
&   1e-7   & 0.517 & -58.081(2) & \multirow{2}{*}{-58.09263(76)} \\
&   1e-8   & 1.041 & -58.085(1) \\
\hline

\multirow{2}{*}{\bf{V}($^4$F)}
&   1e-7   & 0.410 & -71.419(1) & \multirow{2}{*}{-71.44178(59)} \\
&   1e-8   & 0.796 & -71.421(2) \\
\hline

\multirow{2}{*}{\bf{Cr}($^7$S)}
&   1e-7   & 0.347 & -86.618(2) & \multirow{2}{*}{-86.64109(33)} \\
&   1e-8   & 0.883 & -86.625(1) \\
\hline

\multirow{2}{*}{\bf{Mn}($^6$S)}
&   1e-7   & 0.746 & -103.863(2) & \multirow{2}{*}{-103.8919(10)} \\
&   1e-8   & 1.328 & -103.859(2) \\
\hline

\multirow{2}{*}{\bf{Fe}($^5$D)}
&   1e-7   & 0.626 & -123.353(3) & \multirow{2}{*}{-123.38804(93)} \\
&   1e-8   & 1.455 & -123.358(2) \\
\hline

\multirow{2}{*}{\bf{Co}($^4$F)}
&   1e-7   & 0.703 & -145.113(3) & \multirow{2}{*}{-145.1541(10)} \\
&   1e-8   & 1.659 & -145.115(4) \\
\hline

\multirow{2}{*}{\bf{Ni}($^3$F)}
&   1e-7   & 0.456 & -169.335(3) & \multirow{2}{*}{-169.3912(12)} \\
&   1e-8   & 0.921 & -169.345(2) \\
\hline

\multirow{2}{*}{\bf{Cu}($^2$S)}
&   1e-7   & 0.634 & -196.349(3) & \multirow{2}{*}{-196.4038(10)} \\
&   1e-8   & 1.421 & -196.353(3) \\
\hline

\multirow{2}{*}{\bf{Zn}($^1$S)}
&   1e-7   & 0.528 & -226.308(4) & \multirow{2}{*}{-226.3699(18)} \\
&   1e-8   & 1.037 & -226.320(3) \\
\hline

\hline\hline
\end{tabular}
\end{adjustbox}
\end{table}



%% file: kinetic.tex
\section{Kinetic energies}\label{kinetic}

Besides the total energies, we are interested in obtaining accurate kinetic energies for a couple of reasons. 
One reason is that virial theorem does not apply to ECPs due to the modified shape of valence orbitals,
absence of core states and scalar relativity.
Therefore, accurate kinetic energies are unknown. Another important 
point of interest is that kinetic energy provides one possible measure of the degree of the electron density spatial extent/localization. 
This is relevant in QMC calculations where the optimization of the Jastrow factor can and typically does change the density. This can happen in the region close to the nucleus but perhaps even more often in the tail regions. The reason is that the optimized energy is not very sensitive to small 
changes in the tails of the trial function and therefore the optimization method can bias toward regions where the variational gain is the most significant $-$ this is often the region at the largest electronic density. The resulting bias can affect calculations of various quantities, in particular, dipole or higher order moments. 
Values of kinetic energy that differ significantly
from their accurate values provide perhaps the simplest signal of possible bias in the electronic density and imperfect optimization in general.

The kinetic energy of an atom is given by the following expression:
\begin{equation}
    \label{kinetic_expression}
    E_{kin} = \frac{1}{2} \frac{\left\langle \Psi \middle| \nabla^{2} \middle| \Psi \right\rangle}{\left\langle \Psi | \Psi \right\rangle},
\end{equation}
which was obtained from either CISD or FCI calculations with estimations for the CBS limit. 
We estimate the CBS limit energy as follows:
\begin{equation}
    \label{kinetic_CBS}
    E_{kin,\rm CBS}=E_{kin,n}+(E_{kin,n}-E_{kin,n-1}),
\end{equation}
with the corresponding error given by:
\begin{equation}
    \sigma = \frac{{}| E_{kin,n}-E_{kin,n-1}|}{2},
\end{equation}
where n is the largest cardinal number for which the calculation was feasible. 
For TMs, n=5 whereas for all other elements n=6 was used to calculate the CBS limit.

\begin{figure}[htbp!]
\caption{Estimated kinetic energy of ccECP pseudoatoms as a percentage of the total energy, $100E_{kin}/|E_{total}|$.}
\includegraphics[width=\columnwidth]{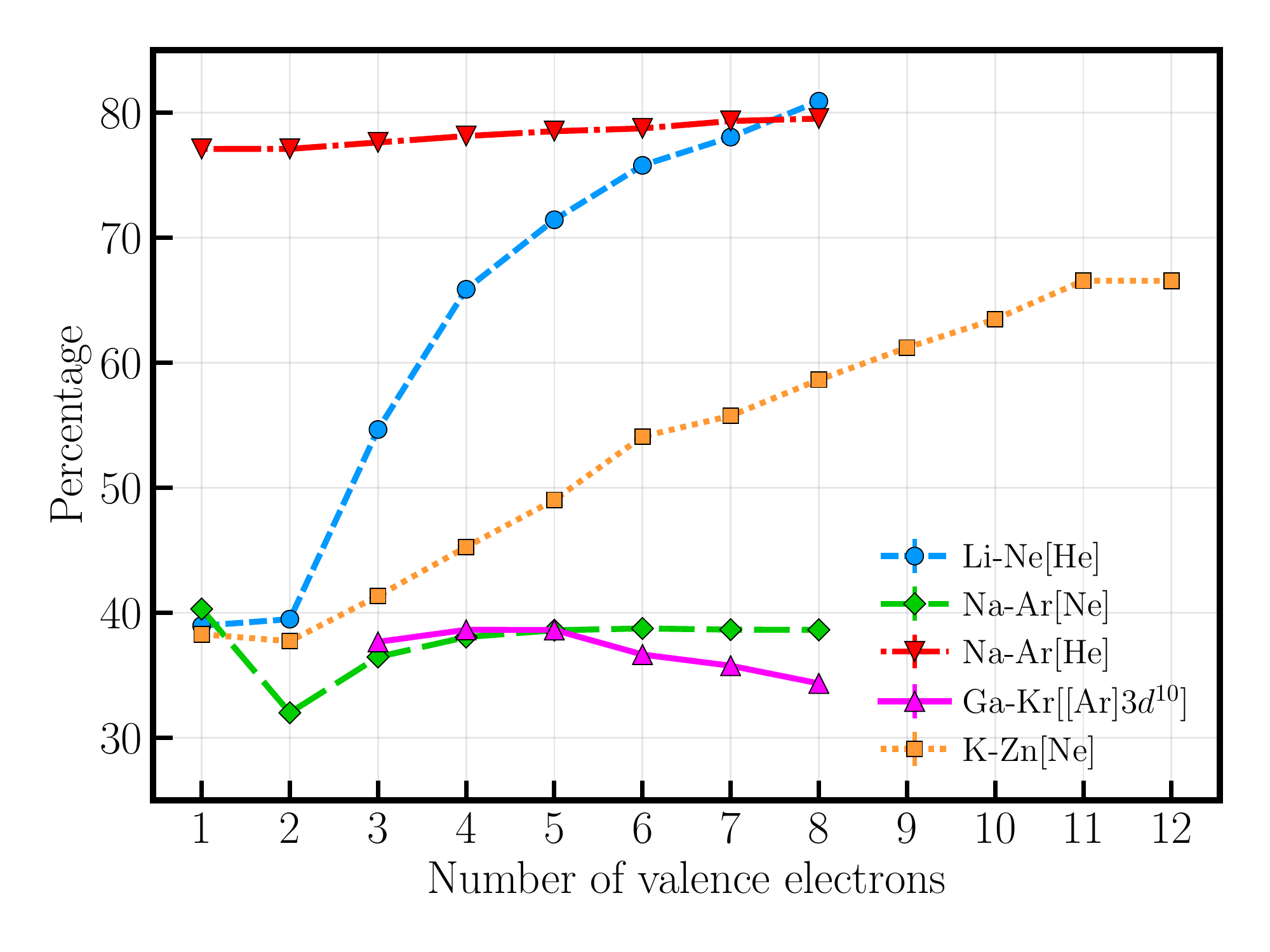}
\label{fig:Kin_ratio}
\centering
\end{figure}

Summaries of all calculated kinetic energies are given tables \ref{tab:1st_ultimate_kin}, \ref{tab:2nd_ultimate_kin}, \ref{tab:2nd-He_ultimate_kin}, \ref{tab:3rd_ultimate_kin}, and \ref{tab:4p_ultimate_kin} which correspond to 1st row ccECP[He], 2nd row ccECP[Ne], 2nd row ccECP[He], K-Zn ccECP[Ne], and Ga-Kr ccECP[[Ar]$3d^{10}$] elements, respectively.
Figure \ref{fig:Kin_ratio} shows these kinetic energies as a ratio to the magnitude of total energy with respect to the number of valence electrons.
Here for K-Zn and 2nd row ccECP[He], semicore electrons were not counted toward the number of valence electrons. 
Having in mind that for AE cases without relativity this ratio is 100\% from the virial theorem, we see that "small" core approximations such as 2nd row with [He] core display a ratio of ($\sim$ 80\%) whereas "large" core cases such as 2nd row with [Ne] core or $4p$ elements with [[Ar]$3d^{10}$] core exhibit much smaller ratios.
The fact that the kinetic to total energy ratio is smaller reflects the smoother nature of pseudoelectronic densities
as a result of regularization of the  Coulomb  singularity and modification of the electron-ion interaction in the core region.
The ratios grow with the number of valence electrons, with the exception of $4p$ elements where the ratio mildly decreases.


\begin{table}[!htbp]
\centering
\caption{Atomic kinetic energies for Li-Ne elements with ccECPs[He]. H and He have only valence electrons and their nuclear Coulomb potential is smoothed out and finite at the nucleus.}
\label{tab:1st_ultimate_kin}
\begin{tabular}{c|cc}
\hline\hline
Atom &  Estimate(Ha)  &  $E_{kin}/|E_{total}|$(\%) \\
\hline
H    &  0.498954(2)   &   99.7909(4)  \\
He   &    2.8855(2)   &    99.372(7)  \\
Li   &  0.07668014(2) &  38.95304(1)  \\
Be   &   0.39884(4)   &    39.480(4)  \\
B    &    1.4292(3)   &     54.65(1)  \\
C    &     3.568(1)   &     65.86(2)  \\
N    &     6.975(1)   &     71.42(1)  \\
O    &    12.036(2)   &     75.77(1)  \\
F    &    18.882(2)   &    78.032(9)  \\
Ne   &    28.342(2)   &    80.902(6)  \\

\hline
\hline
\end{tabular}
\end{table}

\begin{table}[!htbp]
\centering
\caption{Atomic kinetic energies for Na-Ar elements with ccECPs[Ne]. }
\label{tab:2nd_ultimate_kin}
\begin{tabular}{c|cc}
\hline\hline
Atom & Estimate(Ha)  &   $E_{kin}/|E_{total}|$(\%) \\
\hline
Na  & 0.07503(1) & 40.294(5) \\
Mg  & 0.26347(1) & 31.995(2) \\
Al  &  0.7063(1) & 36.454(5) \\
Si  &  1.4314(2) & 38.048(5) \\
P   &  2.4931(3) & 38.594(5) \\
S   &  3.9112(6) & 38.735(6) \\
Cl  &   5.768(1) & 38.642(7) \\
Ar  &  8.1392(8) & 38.626(4) \\
\hline
\hline
\end{tabular}
\end{table}

\begin{table}[!htbp]
\centering
\caption{Atomic kinetic energies for Na-Ar elements with ccECPs[He].}
\label{tab:2nd-He_ultimate_kin}
\begin{tabular}{c|cc}
\hline\hline
Atom & Estimate(Ha)  &   $E_{kin}/|E_{total}|$(\%) \\
\hline
Na  &  36.754(2) &  77.083(4)\\
Mg  &  48.790(2) &  77.091(3)\\
Al  &  63.160(2) &  77.606(2)\\
Si  &  79.719(2) &  78.117(2)\\
P   &  98.693(3) &  78.504(2)\\
S   & 120.013(3) &  78.728(2)\\
Cl  & 144.513(4) &  79.317(2)\\
Ar  & 171.379(5) &  79.514(2)\\
\hline
\hline
\end{tabular}
\end{table}

\begin{table}[!htbp]
    \centering
    \caption{Atomic kinetic energies for K-Zn elements with ccECPs[Ne].}
    \label{tab:3rd_ultimate_kin}
    \begin{adjustbox}{width=\columnwidth,center}
    \begin{tabular}{cc|ccc}
        \hline\hline
        Atom & State & Estimate(Ha)  &   $E_{kin}/|E_{total}|$(\%) \\
        \hline
        K   & ($^2$S) & 10.811(1) & 38.266(4) \\ 
        Ca  & ($^1$S) & 13.856(2) & 37.725(5) \\ 
        Sc  & ($^2$D) & 19.245(4) & 41.336(9) \\ 
        Ti  & ($^3$F) & 26.288(6) &  45.25(1) \\ 
        V   & ($^4$F) & 35.018(7) &  49.02(1) \\ 
        Cr  & ($^7$S) &  46.86(1) &  54.09(1) \\ 
        Mn  & ($^6$S) &  57.91(1) &  55.74(1) \\ 
        Fe  & ($^5$D) &  72.34(1) & 58.628(8) \\ 
        Co  & ($^4$F) &  88.83(1) & 61.197(7) \\ 
        Ni  & ($^3$F) & 107.50(1) & 63.463(6) \\ 
        Ni  & ($^3$D) & 109.15(1) & 64.436(6) \\ 
        Cu  & ($^2$S) & 130.70(1) & 66.547(5) \\ 
        Zn  & ($^1$S) & 150.62(1) & 66.537(4) \\ 
        
        \hline
        \hline
    \end{tabular}
    \end{adjustbox}
\end{table}

\begin{table}[!htbp]
    \centering
    \caption{Atomic kinetic  energies for  Ga-Kr elements with ccECPs[[Ar]$3d^{10}$].}
    \label{tab:4p_ultimate_kin}
    \begin{tabular}{c|ccc}
        \hline\hline
        Atom & Estimate(Ha)  &  $E_{kin}/|E_{total}|$(\%) \\
        \hline
        Ga  &  0.7683(1)  &   37.663(5) \\
        Ge  &  1.4464(3)  &   38.628(8) \\
        As  &  2.3812(5)  &   38.619(8) \\
        Se  &  3.4086(9)  &    36.65(1) \\
        Br  &  4.762(1)   &   35.755(8) \\
        Kr  &  6.344(2)   &    34.34(1) \\
        \hline
        \hline
    \end{tabular}
\end{table}

We now give further details on kinetic energies for each atom such as their values for each basis and method.
Tables \ref{1,ccECP,SA,kin}, \ref{2,ccECP,SA,kin}, \ref{2,ccECP-He,SA,kin}, \ref{3rd-main,ccECP,SA,kin}, and \ref{3,ccECP,SA,kin} give these data for the 1st row, 2nd rows ([Ne] and [He] cores), 3rd row main group and TMs respectively.
For comparison, BFD/eCEPP and STU kinetic energies are also given for selected cases in tables \ref{1,BFD,SA,kin} and \ref{3,STU,SA,kin} respectively.

Note that for 2nd row ccECP[He] and K-Zn ccECP[Ne], FCI calculations were not feasible due to computational limitations and the kinetic energies here are less accurate and estimated from CISD method only. 
Moreover, the kinetic energy for TMs does not show a steady increase in kinetic energy with respect to basis size as opposed to all other cases.


\begin{table*}[htbp!]
\setlength{\tabcolsep}{4pt} 
\centering
\small
\caption{Atomic  kinetic  energies [Ha] for 1st-row elements with ccECPs[He]. aug-cc-pVnZ basis set (cc-pVnZ for Ne).
Values with (*) were not feasible to calculate and represent estimates from the calculated data as described in the text.
}
\label{1,ccECP,SA,kin}
\begin{tabular}{l|l|rrrrr|r}
\hline\hline
Atom & Method &          DZ &          TZ &          QZ &          5Z &          6Z &             CBS \\
\hline

\multirow{1}{*}{\bf{H}}
& ROHF & 0.49895288 & 0.49894527 & 0.49894533 & 0.49894996 & & 0.498954(2) \\
\hline

\multirow{2}{*}{\bf{He}}
& RHF &  2.84397485 & 2.84397376 & 2.84398211 & 2.84397542 & 2.84397523\\
& CISD &  2.85517271 & 2.87995550 & 2.88332835 & 2.88469493 & 2.88512740 & 2.8855(2)\\
\hline

\multirow{1}{*}{\bf{Li}}
& ROHF & 0.076680627 & 0.07668018 & 0.07668022 & 0.076680181 & & 0.07668014(2) \\
\hline

\multirow{2}{*}{\bf{Be}}
& RHF & 0.31533961 & 0.31534286 & 0.31534205 & 0.31534096 & 0.31534214 \\
& CISD & 0.38682436 & 0.39675231 & 0.39818633 & 0.39868344 & 0.39876145 & 0.39884(4) \\
\hline

\multirow{3}{*}{\bf{B}}
& ROHF & 1.31361995 & 1.31362337 & 1.31365191 & 1.31366755 & 1.31367889 \\
& CISD & 1.41475226 & 1.42121447 & 1.42226631 & 1.42282848 & 1.42327868 \\
& FCI  & 1.41542613 & 1.42613533 & 1.42741371 & 1.42804145 & 1.42861308 & 1.4292(3)\\
\hline
 
\multirow{3}{*}{\bf{C}}
& ROHF & 3.43460442 & 3.43464583 & 3.43469478 & 3.43466084 & 3.43467599 \\
& CISD & 3.51446606 & 3.54800763 & 3.55362109 & 3.55425562 & 3.55599662 \\
& FCI  & 3.51956169 & 3.55773200 & 3.56349022 & 3.56417442 & 3.56611352 & 3.568(1) \\
\hline

\multirow{3}{*}{\bf{N}}
& ROHF & 6.82755984 & 6.82752510 & 6.82741059 & 6.82745134 & 6.82750451 \\
& CISD & 6.86959909 & 6.94167021 & 6.95467252 & 6.95730143 & 6.95942600 \\
& FCI  & 6.88381613 & 6.95493764 & 6.96746463 & 6.97040477 & 6.9725(*) & 6.975(1)\\
\hline

\multirow{3}{*}{\bf{O}}
& ROHF & 11.86001287 & 11.84968046 & 11.84346573 & 11.84347378 & 11.84298318 \\
& CISD & 11.84499351 & 11.97842032 & 12.00181762 & 12.00916241 & 12.01384693 \\
& FCI  & 11.87194598 & 11.99674122 & 12.01967110 & 12.0270(*)  & 12.0317(*) & 12.036(2)\\
\hline
 
\multirow{3}{*}{\bf{F}}
& ROHF & 18.63401163 & 18.63769668 & 18.63809632 & 18.63895947 & 18.63905271 &            \\
& CISD & 18.68695319 & 18.81672988 & 18.84119150 & 18.84876588 & 18.85314875 &            \\
& FCI  & 18.71829210 & 18.84140756 & 18.8659(*)  & 18.8735(*)  & 18.8779(*)  &  18.882(2) \\
\hline

\multirow{3}{*}{\bf{Ne}}
& RHF & 28.04250304 & 28.04244770 & 28.04247300 & 28.04248372 & 28.04249187 &            \\
& CISD & 28.08774862 & 28.22056329 & 28.28945965 & 28.30015252 & 28.30424228 &            \\
& FCI  & 28.13283740 & 28.25418488 & 28.3232(*)  & 28.3339(*)  & 28.3380(*)  &  28.342(2) \\

\hline\hline
\end{tabular}
\end{table*}

\begin{table*}[htbp!]
\setlength{\tabcolsep}{4pt} 
\centering
\small
\caption{Atomic  kinetic  energies [Ha] for 2nd-row elements with ccECPs[Ne]. aug-cc-pVnZ basis set (cc-pVnZ for Ar).
Values with (*) were not feasible to calculate and represent estimates from the calculated data as described in the text.
}
\label{2,ccECP,SA,kin}
\begin{tabular}{l|l|rrrrr|r}
\hline\hline
Atom & Method &          DZ &          TZ &          QZ &          5Z &          6Z &             CBS \\
\hline

\multirow{1}{*}{\bf{Na}}
& ROHF & 0.08001484 & 0.07483061 & 0.07499744 & 0.07501148 & & 0.07503(1) \\
\hline

\multirow{2}{*}{\bf{Mg}}
& RHF & 0.23354968 & 0.23199509 & 0.23154500 & 0.23148284 & 0.23146504 \\
& CISD & 0.26548027 & 0.26330003 & 0.26350512 & 0.26349267 & 0.26348253 & 0.26347(1) \\
\hline

\multirow{3}{*}{\bf{Al}}
& ROHF & 0.64168827 & 0.64167743 & 0.64168290 & 0.64168910 & 0.64168131 \\
& CISD & 0.68749744 & 0.69596529 & 0.69837653 & 0.69916991 & 0.69935058 \\
& FCI  & 0.69123162 & 0.70218438 & 0.70489160 & 0.70584552 & 0.70606484 & 0.7063(1) \\
\hline

\multirow{3}{*}{\bf{Si}}
& ROHF & 1.32889986 & 1.32888774 & 1.32888978 & 1.32889540 & 1.32889407 \\
& CISD & 1.38956331 & 1.40885143 & 1.41422020 & 1.41580059 & 1.41619499 \\
& FCI  & 1.39608258 & 1.42180762 & 1.42848773 & 1.43049000 & 1.43096281 & 1.4314(2) \\
\hline

\multirow{3}{*}{\bf{P}}
& ROHF & 2.35490570 & 2.35488943 & 2.35490586 & 2.35489074 & 2.35490457 \\
& CISD & 2.42502036 & 2.45836717 & 2.46828346 & 2.47067375 & 2.47122928 \\
& FCI  & 2.43225546 & 2.47652736 & 2.48959729 & 2.4920(*)  & 2.4926(*) &  2.4931(3) \\
\hline

\multirow{3}{*}{\bf{S}}
& ROHF & 3.70170888 & 3.70169773 & 3.70169403 & 3.70170464 & 3.70170733 \\
& CISD & 3.78836533 & 3.84644311 & 3.86855090 & 3.87327501 & 3.87445746 \\
& FCI  & 3.80049093 & 3.87453937 & 3.90402435 & 3.9088(*)  & 3.9099(*) &  3.9112(6) \\
\hline

\multirow{3}{*}{\bf{Cl}}
& ROHF & 5.49577933 & 5.49575592 & 5.49575015 & 5.49577634 & 5.49575551 \\
& CISD & 5.59949426 & 5.68340388 & 5.71935203 & 5.72725848 & 5.72926018 \\
& FCI  & 5.61423098 & 5.72022380 & 5.756(*)   & 5.764(*)   & 5.766(*)  &  5.768(1) \\
\hline

\multirow{3}{*}{\bf{Ar}}
& RHF & 7.79589481 & 7.79588297 & 7.79588540 & 7.79591613 & 7.79591535 \\
& CISD & 7.94994979 & 8.04531135 & 8.08362249 & 8.08991031 & 8.09141320 \\
& FCI  & 7.95712773 & 8.09128251 & 8.1298(*)  & 8.1361(*)  & 8.1376(*) &  8.1392(8) \\

\hline\hline
\end{tabular}
\end{table*}

\begin{table*}[htbp!]
\setlength{\tabcolsep}{4pt} 
\centering
\small
\caption{Atomic  kinetic  energies [Ha] for 2nd-row elements with ccECPs[He]. cc-pCVnZ basis set.}
\label{2,ccECP-He,SA,kin}
\begin{tabular}{l|l|rrrrr|r}
\hline\hline
Atom & Method &          DZ &          TZ &          QZ &          5Z &          6Z &             CBS \\
\hline

\multirow{2}{*}{\bf{Na}}
& ROHF & 36.52735270 & 36.52739440 & 36.52743838 & 36.52742986 & 36.52742810 &            \\
& CISD & 36.56953299 & 36.71379735 & 36.73737044 & 36.74676260 & 36.75035870 &  36.754(2) \\
\hline

\multirow{2}{*}{\bf{Mg}}
& RHF & 48.55338625 & 48.55353055 & 48.55339130 & 48.55339266 & 48.55340245 &            \\
& CISD & 48.72768236 & 48.76011138 & 48.77236767 & 48.78169464 & 48.78598715 &  48.790(2) \\
\hline

\multirow{2}{*}{\bf{Al}}
& ROHF & 62.89286874 & 62.89280073 & 62.89280702 & 62.89283654 & 62.89282393 &            \\
& CISD & 63.02380551 & 63.11866699 & 63.14118867 & 63.15162101 & 63.15572758 &  63.160(2) \\
\hline

\multirow{2}{*}{\bf{Si}}
& ROHF & 79.41354057 & 79.41348183 & 79.41348239 & 79.41349700 & 79.41352137 &            \\
& CISD & 79.56620941 & 79.67089839 & 79.69816502 & 79.70944994 & 79.71415800 &  79.719(2) \\
\hline

\multirow{2}{*}{\bf{P}}
& ROHF & 98.34763217 & 98.34753767 & 98.34753933 & 98.34759225 & 98.34755827 &            \\
& CISD & 98.51183457 & 98.63625664 & 98.66838143 & 98.68140655 & 98.68736394 &  98.693(3) \\
\hline

\multirow{2}{*}{\bf{S}}
& ROHF & 119.6087101  & 119.6086072  & 119.6086218  & 119.6086240  & 119.6086322  &             \\
& CISD & 119.7876469  & 119.9447924  & 119.9841257  & 119.9997040  & 120.0065278  &  120.013(3) \\
\hline

\multirow{2}{*}{\bf{Cl}}
& ROHF & 144.0465047  & 144.0464073  & 144.0464668  & 144.0465153  & 144.0464488  &             \\
& CISD & 144.2359195  & 144.4298676  & 144.4770235  & 144.4954619  & 144.5040106  &  144.513(4) \\
\hline

\multirow{2}{*}{\bf{Ar}}
& RHF  & 170.8576338  & 170.8574905  & 170.8574457  & 170.8574119  & 170.8575078  &             \\
& CISD & 171.0530054  & 171.2817089  & 171.3365004  & 171.3576565  & 171.3683304  &  171.379(5) \\
\hline

\hline\hline
\end{tabular}
\end{table*}

\begin{table*}[htbp!]
\setlength{\tabcolsep}{4pt} 
\centering
\small
\caption{Atomic  kinetic  energies [Ha] for 3rd-row main group elements with ccECPs. aug-cc-pVnZ basis set (cc-pCVnZ for K, Ca).
Values with (*) were not feasible to calculate and represent estimates from the calculated data as described in the text.
} 
\label{3rd-main,ccECP,SA,kin}
\begin{tabular}{l|l|rrrrr|r}
\hline\hline
Atom & Method &          DZ &          TZ &          QZ &          5Z &          6Z &             CBS \\
\hline

\multirow{2}{*}{\bf{K}}
& ROHF & 10.47154301 & 10.47158948 & 10.47158031 & 10.47154409 & 10.47159990 &            \\
& CISD & 10.67094386 & 10.77344574 & 10.80212489 & 10.80724403 & 10.80933057 &  10.811(1) \\
\hline

\multirow{2}{*}{\bf{Ca}}
& RHF & 13.43105715 & 13.43103412 & 13.43104385 & 13.43103436 & 13.43105075 &            \\
& CISD & 13.68784850 & 13.79155454 & 13.83898979 & 13.84842237 & 13.85206945 &  13.856(2) \\
\hline

\multirow{3}{*}{\bf{Ga}}
& ROHF & 0.71679738 & 0.71676927 & 0.71677262 & 0.71676642 & 0.71676817 &            \\
& CISD & 0.75466761 & 0.76103852 & 0.76303971 & 0.76348371 & 0.76374838 &            \\
& FCI  & 0.75756293 & 0.76513989 & 0.76717098 & 0.76768965 & 0.76797050 &  0.7683(1) \\
\hline

\multirow{3}{*}{\bf{Ge}}
& ROHF & 1.37339649 & 1.37334859 & 1.37336153 & 1.37335122 & 1.37335037 &            \\
& CISD & 1.41820149 & 1.43055699 & 1.43532158 & 1.43615947 & 1.43666937 &            \\
& FCI  & 1.42262827 & 1.43893601 & 1.44429242 & 1.44525212 & 1.44580907 &  1.4464(3) \\
\hline

\multirow{3}{*}{\bf{As}}
& ROHF & 2.29918816 & 2.29920741 & 2.29918168 & 2.29919531 & 2.29918832 &            \\
& CISD & 2.33822054 & 2.35625126 & 2.36544329 & 2.36747993 & 2.36837897 &            \\
& FCI  & 2.34192027 & 2.36650052 & 2.37734101 & 2.37938(*) & 2.38029(*) &  2.3812(5) \\
\hline

\multirow{3}{*}{\bf{Se}}
& ROHF & 3.28128593 & 3.27960996 & 3.27917525 & 3.27916417 & 3.27922877 &            \\
& CISD & 3.32638305 & 3.36556028 & 3.38069776 & 3.38490504 & 3.38664079 &            \\
& FCI  & 3.33284076 & 3.38208369 & 3.40089339 & 3.40512(*) & 3.40687(*) &  3.4086(9) \\
\hline

\multirow{3}{*}{\bf{Br}}
& ROHF & 4.60151346 & 4.60075967 & 4.60066913 & 4.60063168 & 4.60067724 &           \\
& CISD & 4.64604657 & 4.69903055 & 4.72921364 & 4.73656003 & 4.73921506 &           \\
& FCI  & 4.65339016 & 4.71876925 & 4.74907(*) & 4.75645(*) & 4.75912(*) &  4.762(1) \\
\hline

\multirow{3}{*}{\bf{Kr}}
& RHF & 6.15676334 & 6.15659795 & 6.15681251 & 6.15675523 & 6.15681101 &           \\
& CISD & 6.21644385 & 6.27728282 & 6.31615384 & 6.32677745 & 6.33038286 &           \\
& FCI  & 6.22657483 & 6.28751(*) & 6.32644(*) & 6.33708(*) & 6.34069(*) &  6.344(2) \\

\hline\hline
\end{tabular}
\end{table*}

\begin{table*}[htbp!]
\setlength{\tabcolsep}{4pt} 
\centering
\small
\caption{Atomic  kinetic  energies [Ha] for 3rd-row transition elements with ccECPs[Ne]. cc-pCVnZ basis set.}
\label{3,ccECP,SA,kin}
\begin{tabular}{l|l|rrrr|r}
\hline\hline
Atom & Method &          DZ &          TZ &          QZ &          5Z &        CBS \\
\hline

\multirow{2}{*}{\bf{Sc($^2$D)}}
& ROHF & 18.80346523 & 18.78640078 & 18.78741042 & 18.78732902 \\
& CISD & 19.16893469 & 19.20560252 & 19.22999709 & 19.23748738 &  19.245(4) \\
\hline

\multirow{2}{*}{\bf{Ti($^3$F)}}
& ROHF & 25.81869433 & 25.80176707 & 25.80148815 & 25.80150019 \\
& CISD & 26.18810283 & 26.24652940 & 26.26481744 & 26.27662517 &  26.288(6) \\
\hline

\multirow{2}{*}{\bf{V($^4$F)}}
& ROHF & 34.51635252 & 34.51291101 & 34.51252829 & 34.51286327 \\
& CISD & 34.95023504 & 34.99154124 & 34.99093986 & 35.00445192 & 35.018(7) \\
\hline

\multirow{2}{*}{\bf{Cr($^7$S)}}
& ROHF & 46.33993725 & 46.34630254 & 46.34615105 & 46.34601850 \\
& CISD & 46.72814327 & 46.80257628 & 46.81694513 & 46.83905627 &  46.86(1)  \\
\hline

\multirow{2}{*}{\bf{Mn($^6$S)}}
& ROHF & 57.46801485 & 57.45933930 & 57.45978807 & 57.46063830 \\
& CISD & 57.81481658 & 57.86338258 & 57.87035748 & 57.89245175 &  57.91(1) \\
\hline

\multirow{2}{*}{\bf{Fe($^5$D)}}
& ROHF & 71.86877442 & 71.85519345 & 71.85621578 & 71.85688334 \\
& CISD & 72.26058292 & 72.29101858 & 72.29048459 & 72.31381461 &  72.34(1)\\
\hline

\multirow{2}{*}{\bf{Co($^4$F)}}
& ROHF & 88.31280079 & 88.30106436 & 88.30256578 & 88.30292478 \\
& CISD & 88.76131163 & 88.78375033 & 88.77808627 & 88.80332118 &  88.83(1) \\
\hline

\multirow{2}{*}{\bf{Ni($^3$F)}}
& ROHF & 106.9624572  & 106.9563382  & 106.9574403  & 106.9576036  \\
& CISD & 107.4434601  & 107.4653559  & 107.4524209  & 107.4785343  &  107.50(1)\\
\hline

\multirow{2}{*}{\bf{Ni($^3$D)}}
& ROHF & 108.5265119  & 108.5354825  & 108.5336706  & 108.5335871  \\
& CISD & 109.0850457  & 109.1288714  & 109.1054166  & 109.1295428  & 109.15(1)\\
\hline

\multirow{2}{*}{\bf{Cu($^2$S)}}
& ROHF & 130.0368644  & 130.0605199  & 130.0557909  & 130.0552440  \\
& CISD & 130.6059338  & 130.6906066  & 130.6536798  & 130.6786121  &  130.70(1)\\
\hline

\multirow{2}{*}{\bf{Zn($^1$S)}}
& RHF  & 150.1003014  & 150.1018032  & 150.1016129  & 150.1016470  \\
& CISD & 150.6015595  & 150.5955978  & 150.5631498  & 150.5899657  &  150.62(1) \\

\hline\hline
\end{tabular}
\end{table*}

\begin{table*}[htbp!]
\setlength{\tabcolsep}{4pt} 
\centering
\small
\caption{BFD and eCEPP ECPs kinetic energies [Ha]
for selected elements. aug-cc-pVnZ basis set.
Values with (*) were not feasible to calculate and represent estimates from the calculated data as described in the text.
}
\label{1,BFD,SA,kin}
\begin{tabular}{l|l|rrrrr|r}
\hline\hline
Atom & Method &          DZ &          TZ &          QZ &          5Z &          6Z &             CBS \\
\hline

\multicolumn{8}{c}{\textbf{BFD}} \\
\hline

\multirow{3}{*}{\bf{C}}
& ROHF & 3.31094880 & 3.31092724 & 3.31092662 & 3.31094673 & 3.31092544 \\
& CISD & 3.39911868 & 3.43221091 & 3.43739710 & 3.43810930 & 3.43997560 \\
& FCI  & 3.40615656 & 3.44250954 & 3.44799573 & 3.44879783 & 3.45089922 &  3.453(1) \\
\hline

\multirow{3}{*}{\bf{N}}
& ROHF & 6.75184307 & 6.75184179 & 6.75186129 & 6.75186395 & 6.75187214 \\
& CISD & 6.80039239 & 6.87176435 & 6.88452081 & 6.88688084 & 6.88911504 \\
& FCI  & 6.81628005 & 6.88488882 & 6.89760719 & 6.90031892 & 6.902(*)    & 6.905(1)    \\
\hline

\multirow{3}{*}{\bf{O}}
& ROHF & 11.62185052 & 11.61188915 & 11.60608100 & 11.60616593 & 11.60556705 \\
& CISD & 11.61322007 & 11.74923712 & 11.77187918 & 11.77912500 & 11.78416970 \\
& FCI  & 11.64339874 & 11.76677977 & 11.78973117 & 11.797(*) & 11.802(*) &  11.807(3) \\

\hline

\multirow{3}{*}{\bf{Si}}
& ROHF & 1.29870572 & 1.29866559 & 1.29866289 & 1.29866534 & 1.29866035 \\
& CISD & 1.36201200 & 1.38166840 & 1.38671361 & 1.38732416 & 1.38820510 \\
& FCI  & 1.36884674 & 1.39511485 & 1.40164648 & 1.40249072 & 1.40355470 &  1.4046(5) \\
\hline

\multicolumn{8}{c}{\textbf{eCEPP}} \\
\hline
\multirow{3}{*}{\bf{C}}
& ROHF & 3.30805662 & 3.30800792 & 3.30796200 & 3.30800043 & 3.30806491 &            \\
& CISD & 3.39130757 & 3.42251630 & 3.42794440 & 3.42963901 & 3.43019488 &            \\
& FCI  & 3.39705199 & 3.43227414 & 3.43794196 & 3.43986365 & 3.44043734 &  3.4410(3) \\
\hline

\multirow{3}{*}{\bf{N}}
& ROHF & 6.75069887 & 6.75061042 & 6.75062318 & 6.75067900 & 6.75076017 &            \\
& CISD & 6.79195614 & 6.86468788 & 6.87566220 & 6.87910366 & 6.88019007 &            \\
& FCI  & 6.80630062 & 6.87749732 & 6.88840748 & 6.89185(*) & 6.89294(*) &  6.8940(5) \\
\hline

\multirow{3}{*}{\bf{O}}
& ROHF & 11.68788188 & 11.67586754 & 11.67141223 & 11.67139407 & 11.67157750 &            \\
& CISD & 11.66599228 & 11.80881552 & 11.82935374 & 11.83776074 & 11.84126802 &            \\
& FCI  & 11.69516683 & 11.82658946 & 11.84726624 & 11.85568(*) & 11.85919(*) &  11.863(2) \\

\hline\hline
\end{tabular}
\end{table*}

%
%
%
%

\begin{table*}[htbp!]
\setlength{\tabcolsep}{4pt} 
\centering
\small
\caption{Atomic  kinetic  energies [Ha] for selected 3rd-row elements with STU[Ne] ECPs. cc-pCVnZ basis set.}
\label{3,STU,SA,kin}
\begin{tabular}{l|l|rrrr|r}
\hline\hline
Atom & Method &          DZ &          TZ &          QZ &          5Z &        CBS \\
\hline

\multirow{2}{*}{\bf{Fe($^5$D)}}
& ROHF & 72.70128681 & 72.70082006 & 72.70007282 & 72.70231110 &           \\
& CISD & 73.07671018 & 73.12236748 & 73.12007337 & 73.14372524 &  73.17(1) \\
\hline

\multirow{2}{*}{\bf{Co($^4$F)}}
& ROHF & 89.30299210 & 89.30248622 & 89.30181778 & 89.30452148 &           \\
& CISD & 89.73355855 & 89.77189850 & 89.76383273 & 89.78985983 &  89.82(1) \\
\hline

\multirow{2}{*}{\bf{Ni($^3$F)}}
& ROHF & 109.0750991  & 109.0745329  & 109.0738109  & 109.0761206  &            \\
& CISD & 109.5046555  & 109.5419779  & 109.5284339  & 109.5556771  &  109.58(1) \\

\hline\hline
\end{tabular}
\end{table*}

%% file: conclusions.tex
\section{Conclusions}\label{conclusion}

There have been a number of accurate total energy studies for all-electron atoms using a variety of methods such as coupled cluster, DMC with multi-determinant nodes, and FCI, etc (see references \cite{scemama-3d,QMC_AE,troger-C,helium} for selected cases).
However, rigorous examinations of ECP total energies have been rather sporadic and in many cases, only finite basis and limited accuracy methods were employed.
This has been a notable impediment for calculations since many large-scale, solid or bulk calculations employ ECPs/pseudopotentials, especially if heavy atoms such as transition metals are involved.
The missing data on total energies often prevented accurate assessments of methodological errors and usually rather ad hoc estimations or guesses were necessary.
This work provided benchmark data for ccECP atomic ground state energies within current feasibility limits.
Other useful data we provide are values of kinetic energies for atomic ground states which can be rather challenging to obtain and can be used to find balanced QMC optimization of Jastrow factors as mentioned in the text.

There are a couple of conclusions one can draw on the basis of the presented atomic data that will apply to more complicated molecular and solid state systems. In particular, the key consideration is the extent of
the fixed-node DMC errors. At the basic level, the atomic fixed-node error gives a reasonable baseline on the size of the corresponding errors in systems with bonds such as molecules, surfaces or solids.
Considering our previous study, the fixed-node errors tend to \textit{increase} with shorter bond lengths (high electron density) and higher bond multiplicities, while it tends to \textit{decrease} with stretched bond lengths (low electron densities) and single bonded systems \cite{Rasch2014}.
Clearly, more precise assessments are needed and these will be carried out in subsequent studies. 

Additional information that can be inferred from our results is that the correlation energies for other ECP sets can be reasonably estimated 
within about 1\% accuracy using the provided data.
As we explained before, ECP correlation energy is a very mildly varying quantity provided that
the number of channels and semilocal form are the same. 
Mild differences in correlation energies of different ECPs (of the order of 1\%)  can be understood from the degree of smoothness in the core region with a qualitative tendency of slight increase if the resulting density in the core region is larger and smoother (more constant).
This effect is almost fully driven by the valence $s$-channel. 
Our previous papers \cite{2-ccECP,3-ccECP} outline extended discussions and examples of this behavior. 

We believe this study will help to standardize and make much more transparent the properties 
of the newly constructed set of ccECP effective Hamiltonians and also to stimulate further research in this technically demanding but important research area.

%% file: supporting.tex
\section{Supporting Information}

See Supporting Information for further data outlined above and for more details.
The previously provided total and kinetic energies are listed there also for D$_{2h}$ point group. In addition, we tabulate
single-reference FN-DMC energies for each time step at various basis sizes. 
The data for two-configuration FN-DMC for selected cases is provided as well.
The input and output files for this work are shared in Materials Data Facility \cite{MDF}.